\documentclass[a4paper]{article}

\usepackage[margin=1in]{geometry}
\usepackage{placeins}

\usepackage{setspace}

\usepackage{tgpagella}

\usepackage{amsmath}
\usepackage{amssymb}
\usepackage{amsfonts}
\usepackage{bbold}

\usepackage{graphicx}
\usepackage{subcaption}
\usepackage{booktabs}
\usepackage{tabularx}
\usepackage{multirow}

\usepackage{algorithm}
\usepackage[noend]{algpseudocode}

\usepackage{authblk}
\usepackage{siunitx}

\usepackage{hyperref}

\newcommand{\PrintTitle}[1]{%
  \begin{center}
    \LARGE
    \hrule height 0.04in
    \vskip 0.25in
    #1
    \vskip 0.25in
    \hrule height 0.01in
    \vskip 1em
    \normalsize
  \end{center}
}

\newcommand{\PrintAuthors}[1]{%
  \large
  \begin{center}
    #1
  \end{center}
  \vskip 1em
  \normalsize
}

\newcommand{\PrintAuthor}[3]{%
  #1~\textbf{#2}$^{#3}$
}

\newcommand{\PrintAffiliations}[1]{%
  \footnotesize
  \begin{center}
    #1
  \end{center}
  \vskip 1em
  \normalsize
}

\newcommand{\PrintDate}[1]{%
  \large
  \begin{center}
    #1
  \end{center}
  \vskip 4em
  \normalsize
}

\newcommand{\Preprint}{%
  \footnotesize
  \textbf{Preprint} \\
  This version has not been peer-reviewed.
  \normalsize
}

\newcommand{\Citation}[1]{%
  \footnotesize
  \textbf{Citation:} \\
  #1
  \normalsize
}

\newcommand{\CCBY}[1]{%
  \footnotesize
  \copyright #1. This manuscript version is made available under the CC-BY 4.0 license.\\
  \href{https://creativecommons.org/licenses/by/4.0/}{https://creativecommons.org/licenses/by/4.0/}
  \normalsize
}

\usepackage{todonotes}

\begin{document}

\newcommand{\Title}{Neural network scaling and generalization for 3D strain fields in heterogeneous materials}
\newcommand{\Year}{2026}

\PrintTitle{\Title}

\PrintAuthors{
  \PrintAuthor{Lucas}{Fourel}{1\ast},
  \PrintAuthor{Paul}{Devianne}{1},
  \PrintAuthor{Marine}{Debouvry}{1},
  \PrintAuthor{Tristan}{Destruhaut}{1},
  \PrintAuthor{Russell}{Gunn}{2},
  \PrintAuthor{Ashwini}{Gupta}{3},
  \PrintAuthor{Lori}{Graham-Brady}{3},
  \PrintAuthor{Jean-François}{Molinari}{1}
}

\PrintAffiliations{
  $^\ast$lucas.fourel@epfl.ch

  $^1$Civil Engineering Institute, Materials Science and Engineering Institute, École Polytechnique\\Fédérale de Lausanne (EPFL), Station 18, CH-1015 Lausanne, Switzerland

  $^2$Swiss Federal Office of Energy, Switzerland

  $^3$Department of Civil and Systems Engineering, Johns Hopkins University, \\3400 N. Charles Street, Baltimore 21218, MD, USA
}

\PrintDate{
  \Year
}

\begin{abstract}
  Computational homogenization and coupled multiscale simulations of heterogeneous materials require repeated solution of mechanical fields on Representative Volume Elements, which becomes prohibitively expensive for high-resolution three-dimensional microstructures. Machine-learning surrogates can accelerate these computations, yet most existing approaches predict fields for fixed macroscopic loadings and provide limited guidance on how to scale network architectures for high-resolution 3D tensor-field prediction. This work presents a U-Net surrogate trained to predict the full 36-component strain localization tensor field in voxelized concrete mesostructures at resolutions up to 256³. Training localization tensors are generated with an FFT-based linear elastic solver, and an extensive hyperparameter optimization is performed to derive scaling rules and an accuracy-speed trade-off analysis. The results show that only network width and depth need to be scaled with problem complexity, while several conventional U-Net design choices, such as doubling channels after downsampling, are suboptimal for this problem. The final neural network achieves a Mean Absolute Error of 1.118\% on both the validation and the held-out test sets, and an inference speedup of approximately 8.7x over the FFT-based solver on an Nvidia H100 GPU. Generalization experiments reveal transfer across aggregate volume fractions and highlight the importance of training on diverse morphologies. To safeguard reliability, a Maximum Mean Discrepancy metric computed on internal feature maps is proposed for detecting out-of-distribution microstructures without costly reference computations.
\end{abstract}

\vfill

\Preprint

\Citation{
  L. Fourel, P. Devianne, M. Debouvry, T. Destruhaut, R. Gunn, A. Gupta, L. Graham-Brady, J.-F. Molinari. \Title. \Year.
}

\CCBY{\Year}

\clearpage

\clearpage

\section{Introduction}

In engineering applications, solid materials are generally considered homogeneous at the macroscopic scale, although most of them are heterogeneous at a finer scale. For instance, concrete is composed of aggregates embedded in a cementitious matrix and these two phases possess different mechanical properties, leading to non-uniform stress and strain fields that govern macroscopic performance and durability. The same principle applies to fiber-reinforced composites, and polycrystalline materials.

Computational methods such as the Finite Element Method (FEM) and Fast Fourier Transform (FFT) based solvers \cite{moulinec1994,moulinec1998} can be used to compute stress and strain fields in heterogeneous materials. These fields are generally used for homogenization or coupled multiscale simulation purposes. In one-way homogenization, the objective is to derive a macroscopic effective material behavior that will be assigned to the equivalent homogeneous material in macroscopic models. In coupled multiscale approaches, such as FE² \cite{feyel2000} and FE-FFT \cite{spahn2014}, there is no macroscopic constitutive law, and the material behavior is evaluated at each macroscopic integration point by computing and averaging the stress in a heterogeneous Representative Volume Element (RVE) subjected to the macroscopic strain. The multiscale methods can require a very large number of RVE computations depending on the size of the macroscopic mesh and the number of iterations necessary to reach convergence between the two scales (in staggered solving schemes). Unfortunately, for microstructures (or mesostructures) with complex geometries, the computational cost can be very large due to the high resolution required to resolve the fine details. Due to the numerous RVE computations and the potentially high computational cost of each of them, multiscale simulation approaches often lead to infeasible computational times even though these approaches have a tremendous potential to improve the accuracy of simulations by bridging multiple scales, which helps understand and predict complex physical phenomena involving the interaction between a macroscopic problem and microscopic (or mesoscopic) features.

In recent years, Machine Learning (ML) has emerged as a promising paradigm for accelerating the computations of stress and strain fields in heterogeneous materials \cite{bishara2023,wang2024a}. Instead of computing these fields on-the-fly for each new microstructure or new loading condition, a dataset is precomputed using traditional numerical methods (generally FEM or FFT-based solvers), an ML model is trained on this dataset, and then used as a surrogate for the numerical model. The computation using the surrogate is often faster than the original at the expense of an additional approximation error. This approach is particularly beneficial for multiscale simulations requiring numerous RVE computations for which the computational cost is critical, although ML surrogate models can be used on their own, to evaluate stress or strain fields in new microstructures and eventually to derive macroscopic effective behaviors. One important limitation is that the generalization outside the training data distribution is not guaranteed.

In 2015, Liu et al. \cite{liu2015} trained multiple ML models including decision trees, random forests and support vector regression models to predict linear elastic strain fields in 3D microstructures and found that random forest was the best approach for their configuration, with a Mean Absolute Error (MAE) of \SI{13}{\percent}. In 2019, Yang et al. \cite{yang2019} discovered that neural networks, and particularly Convolutional Neural Networks (CNNs) were even better than random forest, reaching \SI{3}{\percent} MAE on the same task.
Khanolkar et al. \cite{khanolkar2021} showed CNN capabilities on a wide range of microstructures.
A CNN variation, the U-Net architecture \cite{ronneberger2015}, first developed for medical image segmentation purposes, was found to be very efficient at predicting stress and strain fields in composites \cite{raj2021,bhaduri2022,gupta2023,saha2024} and polycrystalline microstructures \cite{mianroodi2021,khorrami2023,langcaster2024}.
Yang et al. \cite{yang2021} used a generative adversarial training approach on a U-Net for similar tasks, and Monteiro Fernandes et al. \cite{monteirofernandes2025} compared training based on a physics-informed loss function (minimization of the stress divergence) and training based on simulation data.
Other studies have trained U-Nets on more complex microstructures with varying morphologies \cite{croom2022,shokrollahi2023} and volume fractions \cite{yao2025}. In addition to synthetic microstructures, CNNs are also able to predict stress fields in microstructures measured by X-ray tomography \cite{sun2024}.
Researchers have successfully trained U-Nets to predict not only strain fields but also fracture patterns \cite{sepasdar2022,chen2023a}.
Alternative architectures, including graph neural networks \cite{maurizi2022,yacouti2025}, Fourier neural operators \cite{rashid2022}, difference-based networks \cite{feng2021}, and hybrid convolution-attention U-Nets \cite{li2025,wei2026} have also shown good capabilities.

Most existing ML surrogates for heterogeneous materials predict microscopic stress or strain fields for one or a few fixed macroscopic loadings, which limits their usefulness in multiscale frameworks where arbitrary loading paths must be evaluated. Predicting the full strain localization tensor instead provides a more general representation: a single network forward pass yields a microstructure-specific linear operator from which strain and stress fields can be reconstructed for any macroscopic strain in the linear elastic regime. The same concept extends to tangent operators in nonlinear problems, provided the prediction is conditioned on the current strain state. This tensorial formulation is therefore both physically grounded and computationally advantageous. Furthermore, while U-Nets have repeatedly been shown to perform well on field-prediction tasks, the literature lacks a systematic investigation of how their architecture should be scaled for high-resolution three-dimensional tensor fields, and how hyperparameters should be chosen to balance predictive accuracy against the computational speed required in multiscale simulations. The present work addresses these gaps by (1) training a parametric 3D U-Net to predict the complete \num{36}-component strain localization tensor in $64^3$ to $256^3$ concrete mesostructures, (2) performing an extensive hyperparameter optimization to derive robust scaling rules and an accuracy-speed trade-off analysis, and (3) assessing the generalization limits of such surrogates when the training microstructural distribution is varied. The resulting model is intended as a fast, data-driven substitute for RVE solvers in coupled multiscale simulations, while highlighting the practical trade-offs that must be navigated when deploying machine-learning surrogates in computational mechanics.

\section{Methods}
\label{sec:methods}

\subsection{Task}
\label{sec:task}

The neural network is trained to predict the strain localization tensor field $\mathbb{A}$ based on the microstructure. This tensor governs the mapping from the macroscopic strain $\bar{\varepsilon}$ to the microscopic strain $\varepsilon$:
\begin{equation}
  \varepsilon \left( x \right) = \mathbb{A} \left( x \right) : \bar{\varepsilon}
  \label{eq:localization}
\end{equation}
Consequently, the microscopic stress $\sigma$ is obtained via:
\begin{equation}
  \sigma \left( x \right) = \mathbb{C} \left( x \right) : \mathbb{A} \left( x \right) : \bar{\varepsilon}
  \label{eq:localization-stress}
\end{equation}
Determining the localization tensor enables homogenization and provides access to local strain and stress fields. While $\mathbb{A}$ is constant for linear elastic materials, the same construction yields a tangent operator in nonlinear regimes, where it depends on the strain state or material history and maps increments of macroscopic strain to increments of microscopic strain. The nonlinear case is, however, substantially more demanding: the operator must be conditioned on the current state, and it must be identified from perturbed rather than unit loadings. This study focuses exclusively on the linear elastic regime.

The localization tensor $\mathbb{A}$ exhibits the following properties:
\begin{itemize}
  \item $\mathbb{A}$ is a fourth-order tensor field dependent on the spatial position $x$.
  \item Due to the symmetry of $\bar{\varepsilon}$ and $\varepsilon$, $\mathbb{A}$ possesses minor symmetries. It can thus be represented as a $6 \times 6$ matrix field using Voigt-Mandel notation.
  \item Unlike the local stiffness tensor $\mathbb{C}$, $\mathbb{A}$ generally lacks major symmetry; therefore, it retains \num{36} independent components rather than reducing to \num{21}.
  \item The volume average of $\mathbb{A} \left( x \right)$ equals the identity tensor $I$.
  \item The volume average of $\mathbb{C} \left( x \right) : \mathbb{A} \left( x \right)$ yields the effective stiffness tensor $\bar{\mathbb{C}}$, relating macroscopic stress to strain: $\bar{\sigma} = \bar{\mathbb{C}} : \bar{\varepsilon}$.
\end{itemize}

Given the \num{36} components of the localization tensor, the microscopic strain and stress fields (both \num{6}-component fields) can be derived for arbitrary macroscopic loading using \autoref{eq:localization} or \autoref{eq:localization-stress}. The neural network input is a scalar phase map of the microstructure, and the output is the corresponding \num{36}-component tensor field.

To train and evaluate the surrogate, we first generate a reference dataset of microstructures and corresponding localization tensors.

\subsection{Data}
\label{sec:data}

This study employs synthetic data generated via computational algorithms. However, the proposed methodology is directly applicable to experimental data obtained from real specimens.

\subsubsection{Microstructure generation}
\label{sec:data-in}

We investigate concrete mesostructures, though the term microstructure is also used hereafter for generality. These mesostructures comprise two phases: cement paste (matrix) and aggregates (particles).

\begin{figure}[ht]
  \centering
  \begin{subfigure}[b]{0.32\textwidth}
    \includegraphics[width=\textwidth]{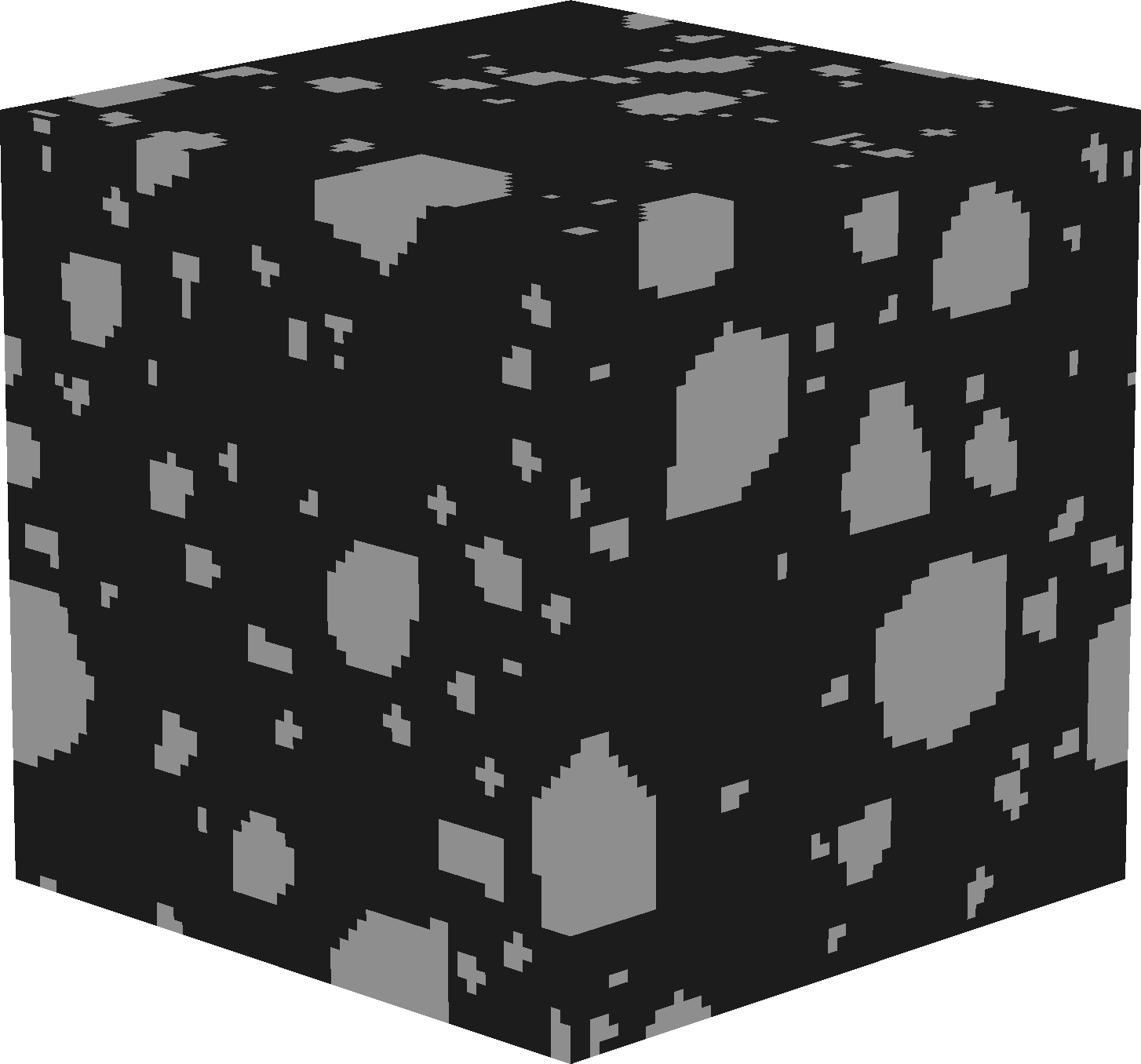}
    \caption{$64^3$}
    \label{fig:r64}
  \end{subfigure}
  \hfill
  \begin{subfigure}[b]{0.32\textwidth}
    \includegraphics[width=\textwidth]{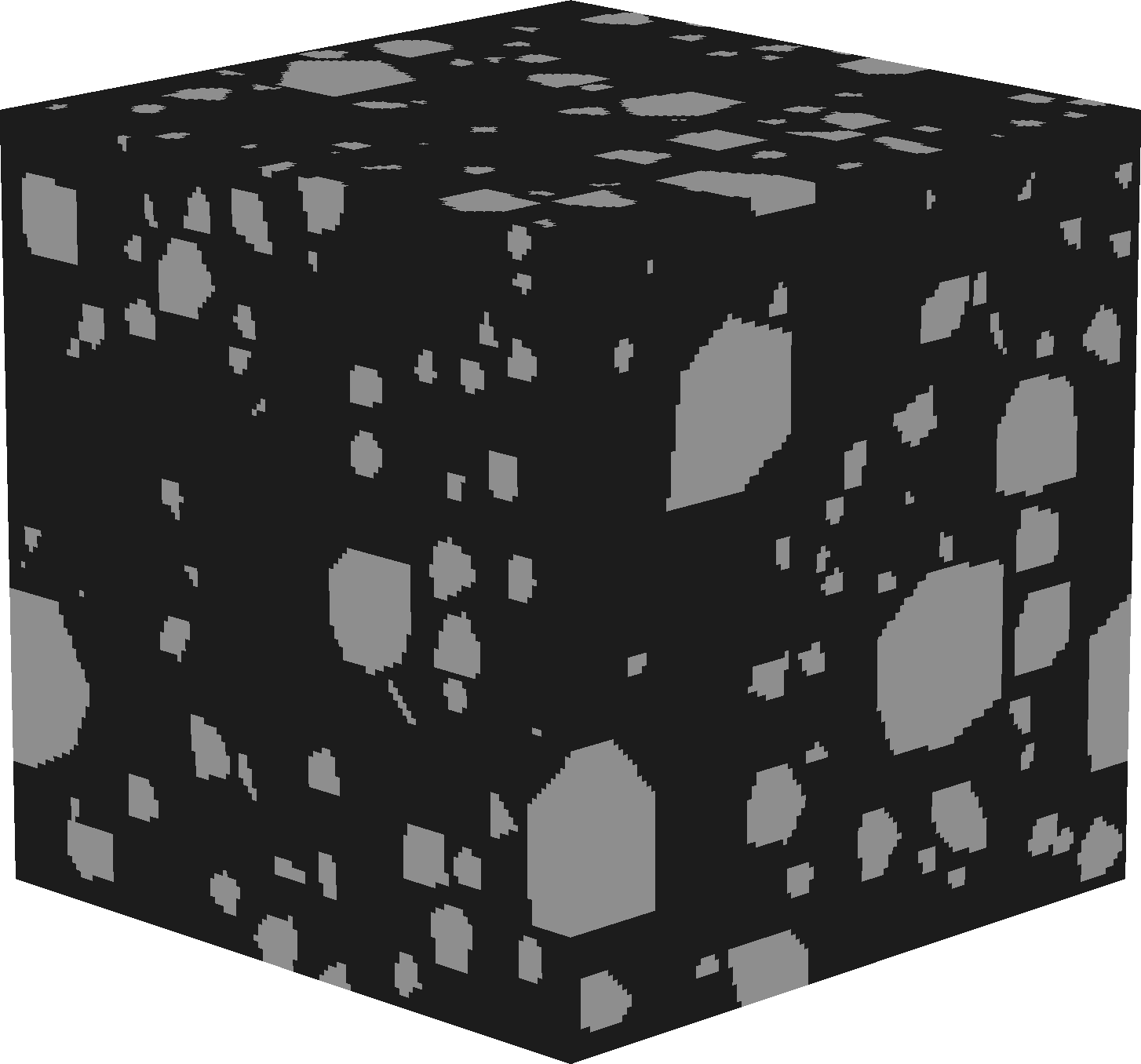}
    \caption{$128^3$}
    \label{fig:r128}
  \end{subfigure}
  \hfill
  \begin{subfigure}[b]{0.32\textwidth}
    \includegraphics[width=\textwidth]{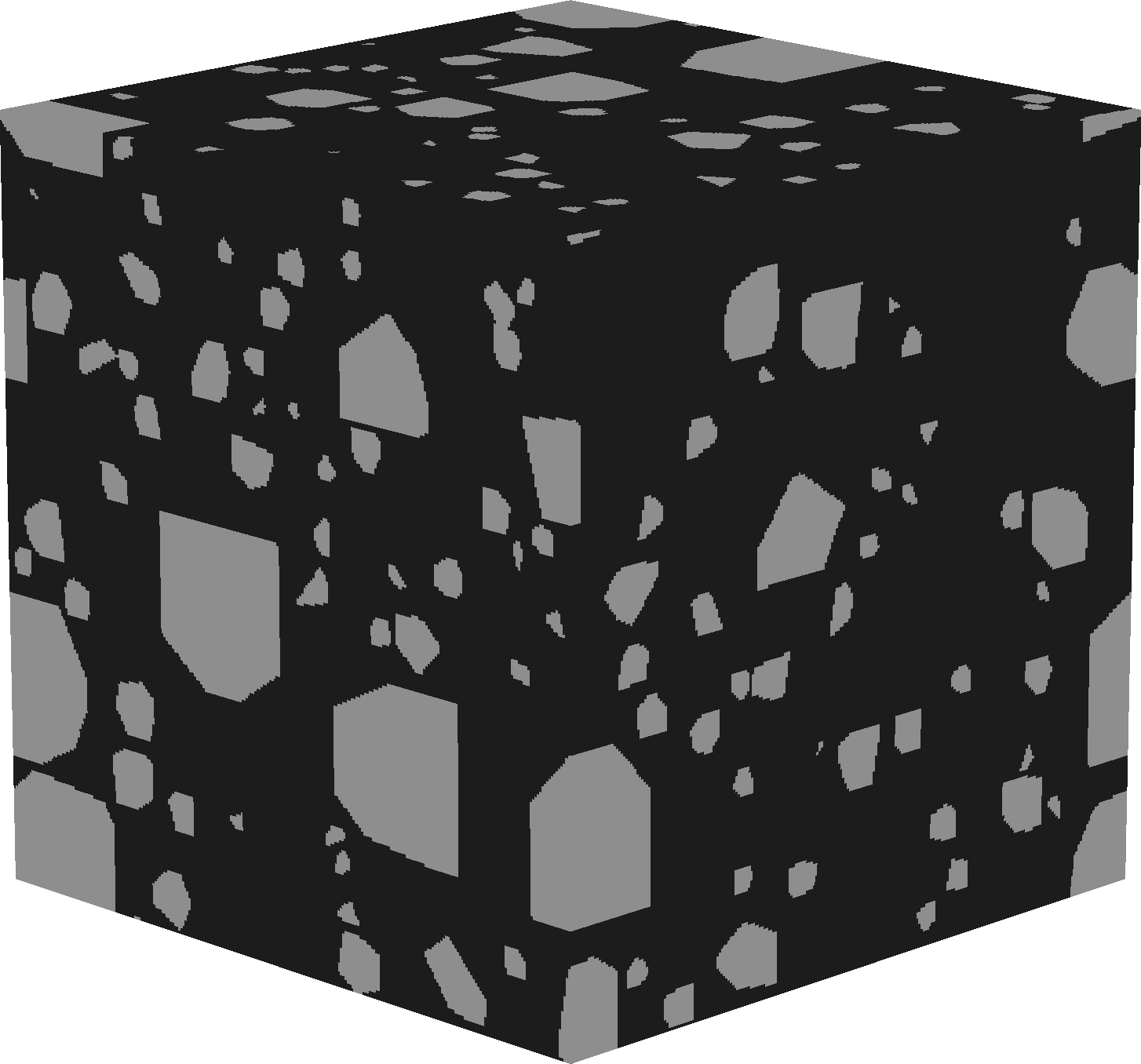}
    \caption{$256^3$}
    \label{fig:r256}
  \end{subfigure}
  \caption{Concrete mesostructures generated at three different resolutions. Samples correspond to an aggregate volume fraction (VF) of \SI{20}{\percent} and a shape ratio of \SI{100}{\percent}.}
  \label{fig:resolutions}
\end{figure}

\begin{figure}[ht]
  \centering
  \begin{subfigure}[b]{0.245\textwidth}
    \includegraphics[width=\textwidth]{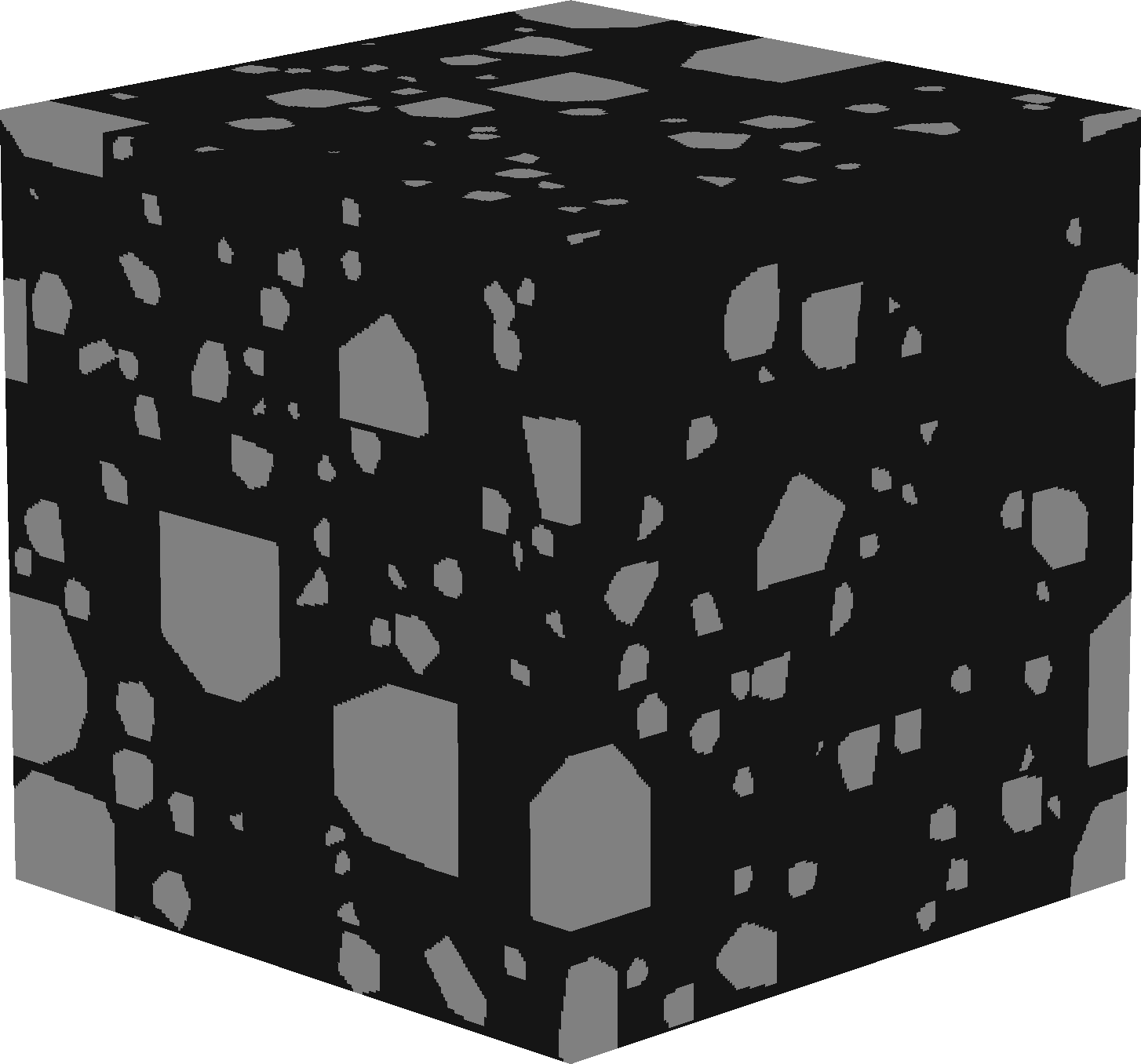}
    \caption{VF: \SI{20}{\percent}, Ratio: \SI{100}{\percent}}
    \label{fig:f20-s100}
  \end{subfigure}
  \hfill
  \begin{subfigure}[b]{0.245\textwidth}
    \includegraphics[width=\textwidth]{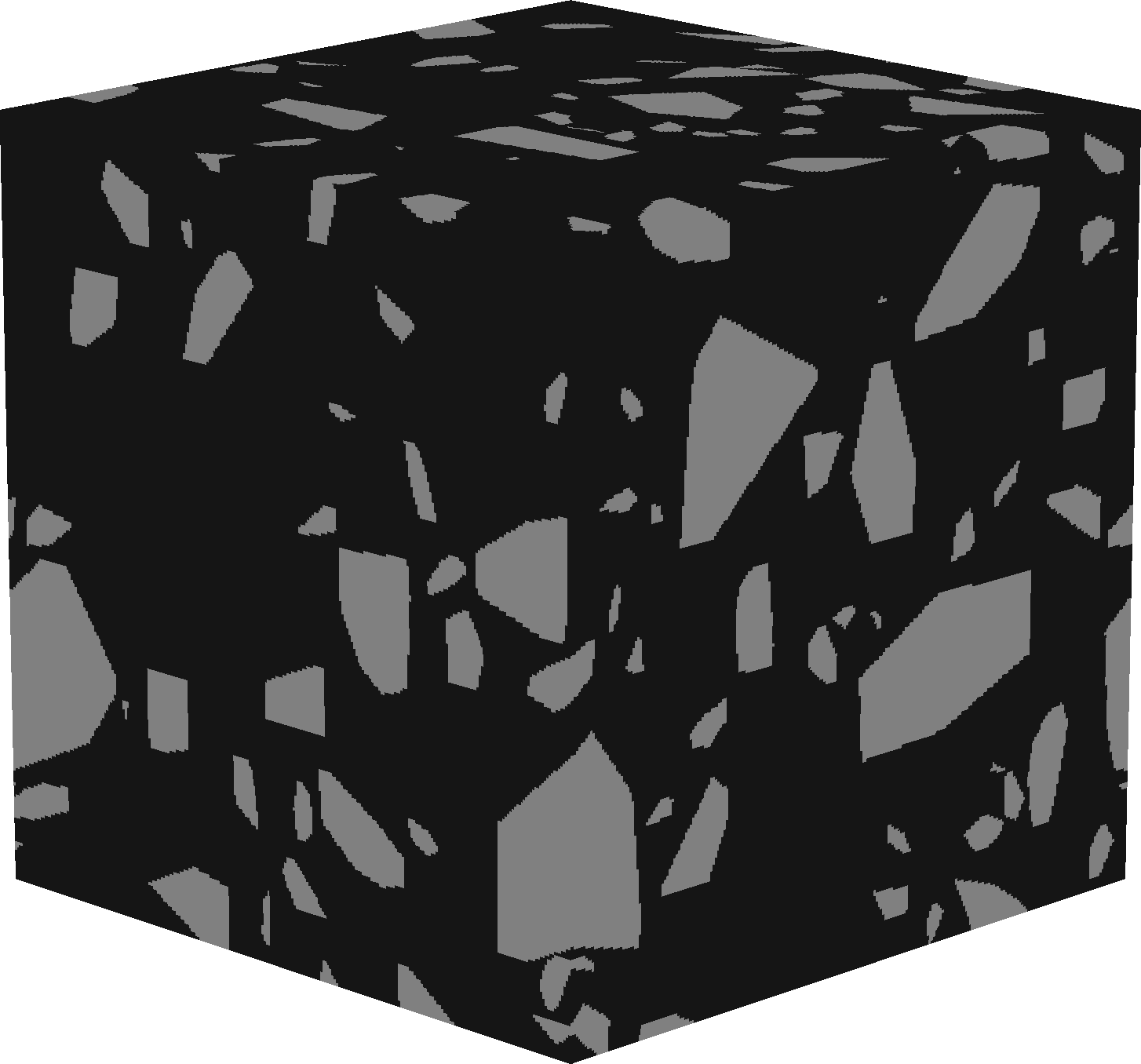}
    \caption{VF: \SI{20}{\percent}, Ratio: \SI{150}{\percent}}
    \label{fig:f20-s150}
  \end{subfigure}
  \hfill
  \begin{subfigure}[b]{0.245\textwidth}
    \includegraphics[width=\textwidth]{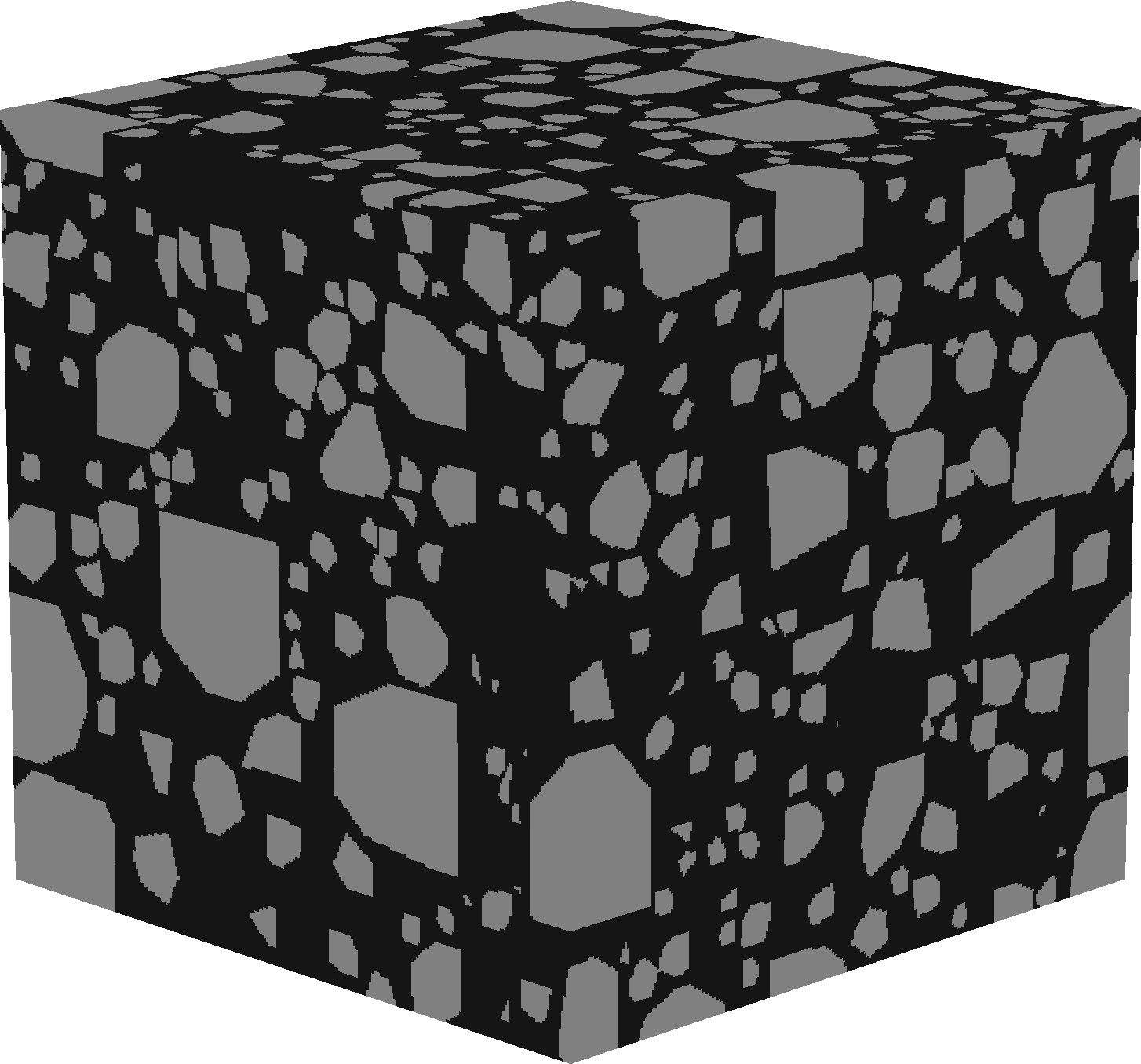}
    \caption{VF: \SI{40}{\percent}, Ratio: \SI{100}{\percent}}
    \label{fig:f40-s100}
  \end{subfigure}
  \hfill
  \begin{subfigure}[b]{0.245\textwidth}
    \includegraphics[width=1\textwidth]{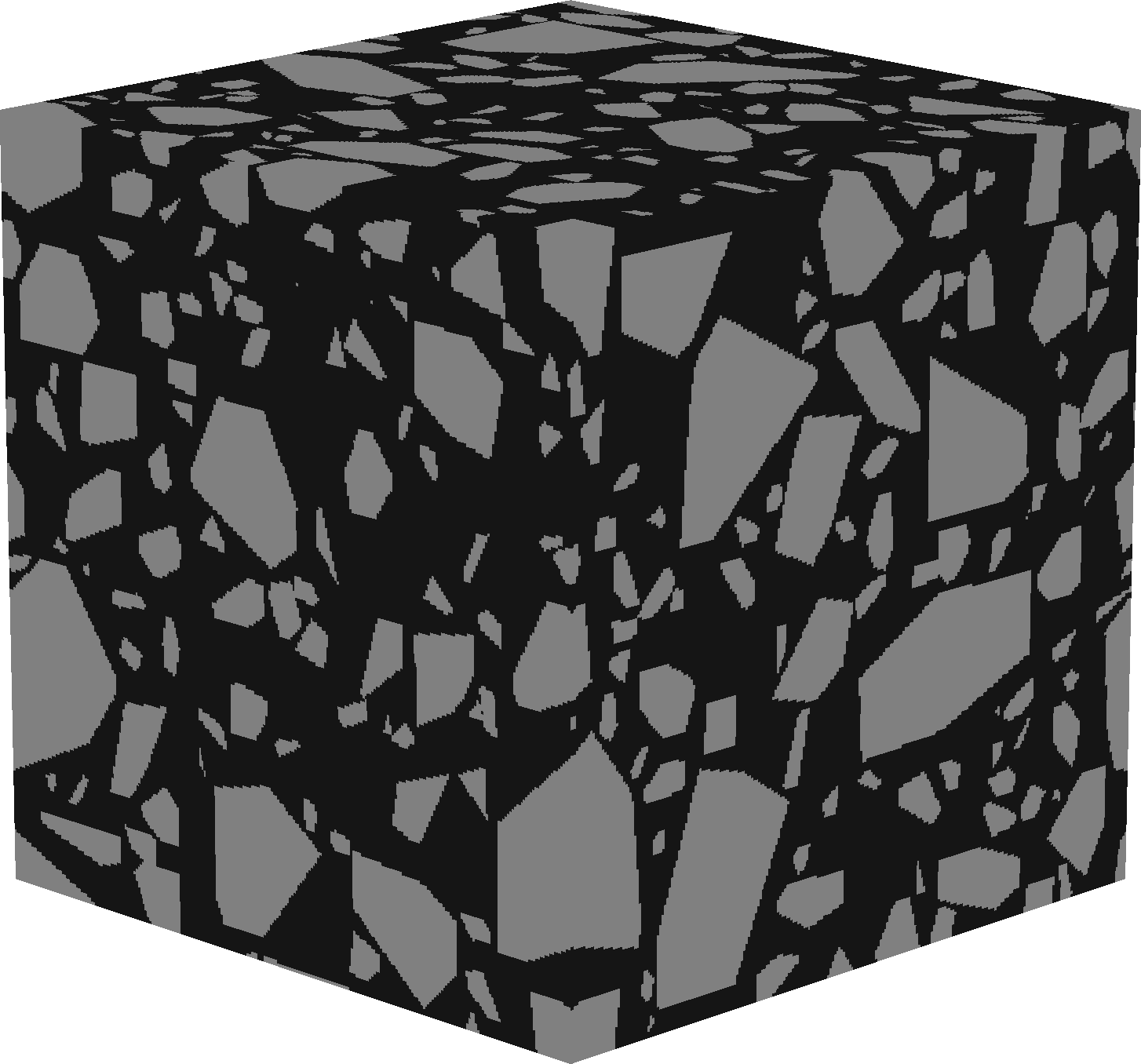}
    \caption{VF: \SI{40}{\percent}, Ratio: \SI{150}{\percent}}
    \label{fig:f40-s150}
  \end{subfigure}
  \caption{Mesostructures generated with varying aggregate volume fractions (VF) and shape ratios. Samples correspond to resolution $256^3$.}
  \label{fig:morphologies}
\end{figure}

Periodic voxel-based mesostructures were generated to ensure compatibility with the FFT-based solver used to compute the strain localization tensor (see \autoref{sec:data-out}). Generation was performed using the open-source package PyCMG \cite{holla2021}, which implements a random sequential adsorption algorithm. We varied the resolution, aggregate volume fraction (VF), and aggregate morphology, as illustrated in \autoref{fig:resolutions} and \autoref{fig:morphologies}. All mesostructures share an aggregate size distribution following a power law with exponent \num{0.5} (Fuller curve), with equivalent diameters ranging from \SI{5}{\percent} to \SI{25}{\percent} of the domain size. The problem is formulated in dimensionless units by normalizing all characteristic lengths with respect to the domain size.

Resolutions were selected as powers of \num{2} to facilitate downsampling: $64^3$, $128^3$, and $256^3$. Aggregate volume fractions were set to \SI{20}{\percent} and \SI{40}{\percent}. These values are lower than typical concrete fractions to account for fine aggregates (smaller than \SI{5}{\percent} of the domain size) that cannot be resolved at the lowest resolution and are effectively homogenized into the cement matrix. The aggregate shape parameter defines the aspect ratio (ratio of largest to smallest axes). A ratio of \SI{150}{\percent} yields elongated aggregates, whereas \SI{100}{\percent} corresponds to equiaxed polyhedra (see \autoref{fig:morphologies}).
For each resolution and parameter combination, \num{256} random samples were generated.

\FloatBarrier

\FloatBarrier

With the mesostructures in place, the next step is to compute the associated strain localization tensors.

\subsubsection{Strain computation}
\label{sec:data-out}

To generate the output examples for the neural network training, the strain localization tensor field $\mathbb{A}(x)$ was computed for each generated microstructure using an FFT-based solver \cite{moulinec1994,moulinec1998}. This approach leverages the periodicity of the generated voxel grids to solve the Lippmann-Schwinger integral equation associated with the linear elastic equilibrium problem efficiently.

The local stiffness tensor $\mathbb{C}(x)$ is defined piecewise based on the phase map. The cement matrix is assigned an isotropic stiffness tensor $\mathbb{C}_m$, while the aggregates are assigned a stiffer isotropic tensor $\mathbb{C}_a$. In dimensionless units, the Young's modulus of the matrix is set to $E_m = 1$, and the aggregate modulus is set to $E_a = 5$, representing a typical contrast found in concrete mesostructures. The Poisson's ratio is fixed at $\nu = 0.2$ for both phases.

To fully reconstruct the $6 \times 6$ localization matrix at every voxel, the unit cell is subjected to six linearly independent macroscopic strain loadings $\bar{\varepsilon}^{(k)}$ ($k=1, \dots, 6$), corresponding to the basis vectors in Voigt-Mandel notation:
\begin{equation}
  \bar{\varepsilon}^{(1)} =
  \begin{bmatrix} 1 \\ 0 \\ 0 \\ 0 \\ 0 \\ 0
  \end{bmatrix}, \dots, \bar{\varepsilon}^{(6)} =
  \begin{bmatrix} 0 \\ 0 \\ 0 \\ 0 \\ 0 \\ 1
  \end{bmatrix}.
\end{equation}
For each loading case $k$, the solver computes the resulting microscopic strain field $\varepsilon^{(k)}(x)$. The $k$-th column of the localization matrix $\mathbb{A}(x)$ in Voigt-Mandel notation is then directly identified as the vector $\varepsilon^{(k)}(x)$. By concatenating the results from all six loadings, we obtain the complete tensor field $\mathbb{A}(x) \in \mathbb{R}^{36}$ for each sample.

The FFT-based solver used in this work relies on the original Lippmann-Schwinger strain formulation. Spatial differential operators are discretized using a forward-backward finite difference scheme, reducing the typical FFT-based solver artifacts.
The conjugate gradient method is employed with a convergence criterion based on the equilibrium residual. The iterative process is terminated when the $L_2$-norm of the residual falls below $10^{-6}$. All simulations were performed on a high-performance computing cluster, using Nvidia H100 SMX5 \SI{94}{\giga\byte} GPUs.

\begin{figure}[ht]
  \centering
  \includegraphics[width=1\textwidth]{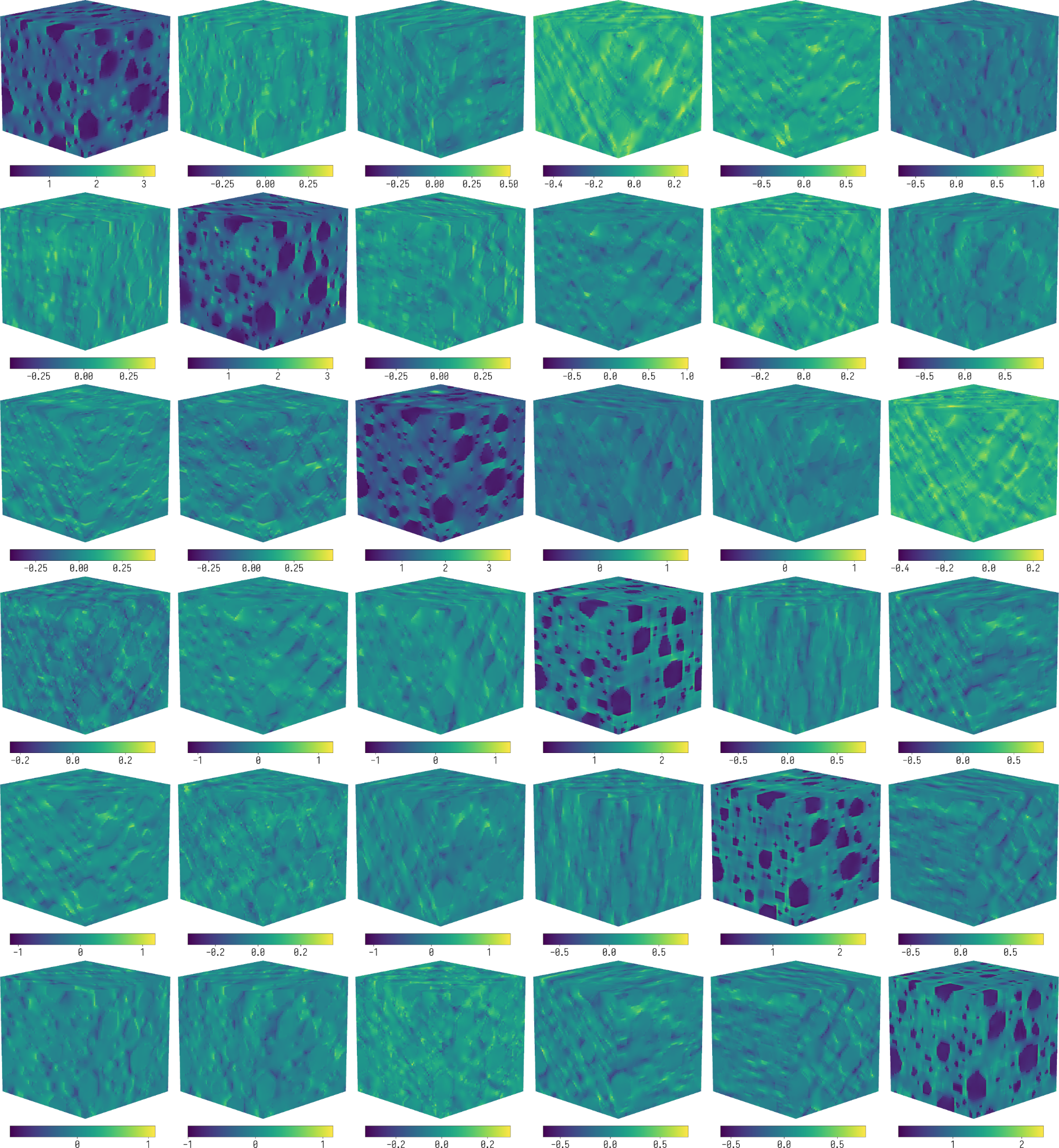}
  \caption{Visualization of the strain localization tensor field for a representative mesostructure ($64^3$ voxels, \SI{20}{\percent} aggregate volume fraction, shape ratio \num{1.0}). Each column gives the microscopic strain induced by a unit macroscopic loading, with rows denoting individual strain components. The tensor is dimensionless, mapping the macroscopic strain $\bar{\varepsilon}$ to the microscopic strain $\varepsilon(x)$.}
  \label{fig:localization}
\end{figure}

A representative strain localization tensor field is visualized in \autoref{fig:localization}. The \num{36} independent components are organized into a $6 \times 6$ matrix using Voigt-Mandel notation. Consistent with the formulation in \autoref{sec:task}, each column corresponds to the microscopic strain response induced by a specific unit macroscopic loading, while the rows represent the individual strain components.

\FloatBarrier

\subsubsection{Data splitting}
\label{sec:data-splitting}

The generated dataset comprising \num{256} samples for each configuration (resolution, volume fraction, and shape ratio) was partitioned into training, validation, and test sets to ensure robust model evaluation and prevent data leakage. A stratified splitting strategy was employed to maintain a consistent distribution of microstructural parameters across all subsets. Specifically, \SI{75}{\percent} of the data (192 samples per configuration) was allocated for training the neural network weights. The validation set, comprising \SI{12.5}{\percent} (32 samples), was used for monitoring convergence and hyperparameter tuning. The remaining \SI{12.5}{\percent} (32 samples) constituted the held-out test set, which was strictly reserved for the final assessment of the model's generalization capabilities on unseen microstructures. This test set was evaluated only once, after training and hyperparameter selection were complete, and the resulting test error is reported together with the validation error in \autoref{sec:res-final}. This splitting protocol was applied independently to each group of samples defined by its specific resolution and morphological parameters to guarantee that model performance is evaluated fairly across the entire design space.

We now describe the neural network architecture used to map the microstructure to the localization tensor.

\subsection{Architecture}
\label{sec:architecture}

Given the voxel-based representation of the microstructures and the necessity for translation equivariance, a Convolutional Neural Network (CNN) was selected as the backbone \cite{lecun1998}. Specifically, we employ the U-Net architecture \cite{ronneberger2015}, which consists of a contracting path (encoder) and an expanding path (decoder). The encoder comprises multiple levels, each containing convolutional layers that capture features across varying frequencies: high-frequency details at higher resolutions and low-frequency features in downsampled representations. This symmetric structure is particularly well-suited for image-to-image regression tasks where input and output share identical spatial dimensions.

Spatial downsampling is performed via $2 \times 2 \times 2$ convolutions with stride $2$. The decoder mirrors this process using transposed convolutions to progressively restore resolution. Crucially, skip connections shuttle feature maps from each encoder level to its counterpart in the decoder, concatenating them before further processing so that high-resolution spatial information directly informs the final prediction.

\begin{figure}[ht]
  \centering
  \includegraphics[width=1\textwidth]{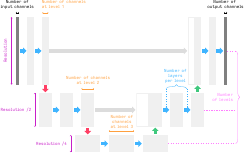}
  \caption{Schematic representation of the parametric U-Net architecture. The diagram illustrates the encoder-decoder structure with multiple levels. Blue arrows indicate convolutional operations, red arrows denote downsampling, green arrows denote upsampling, and gray arrows represent skip connections.}
  \label{fig:architecture-diagram}
\end{figure}

We implement a parametric U-Net variant designed to facilitate systematic hyperparameter exploration. The architecture exposes several degrees of freedom that jointly control its capacity and computational cost:
\begin{itemize}
  \item \textbf{Number of channels} (integer): The base network width, defining the number of feature maps in the initial encoder level. The channels can be seen as the components in the case of tensor fields.
  \item \textbf{Number of levels} (integer): The depth of the spatial hierarchy, i.e., the number of distinct resolution scales.
  \item \textbf{Number of layers per level} (integer): The number of sequential convolutional blocks allocated to each level.
  \item \textbf{Kernel size} (integer): The spatial extent $K$ of the $K \times K \times K$ convolutional windows.
  \item \textbf{Separated convolution} (boolean): Whether standard convolutions are factorized into separate spatial and channel-wise operations.
  \item \textbf{Constant channels} (boolean): Whether the channel count remains fixed or doubles upon each downsampling step.
  \item \textbf{Bias} (boolean): Whether bias terms are included in the linear transformations.
  \item \textbf{Skip connections} (categorical): The topology governing how encoder features are routed to the decoder. Options include direct concatenation, no skip connections, or transforming encoder features via an additional convolution operation before passing them to the decoder at constant resolution.
  \item \textbf{Normalization method} (categorical): The normalization technique applied after convolution (e.g., batch, instance, disabled).
  \item \textbf{Activation function} (categorical): The nonlinear activation used in hidden layers (e.g., LeakyReLU, Sigmoid, or SiLU).
\end{itemize}
\autoref{fig:architecture-diagram} provides a schematic overview of this U-Net, illustrating how the hyperparameters govern the flow from input phase map to output tensor field.

\FloatBarrier

\begin{figure}[ht]
  \centering
  \includegraphics[width=1\textwidth]{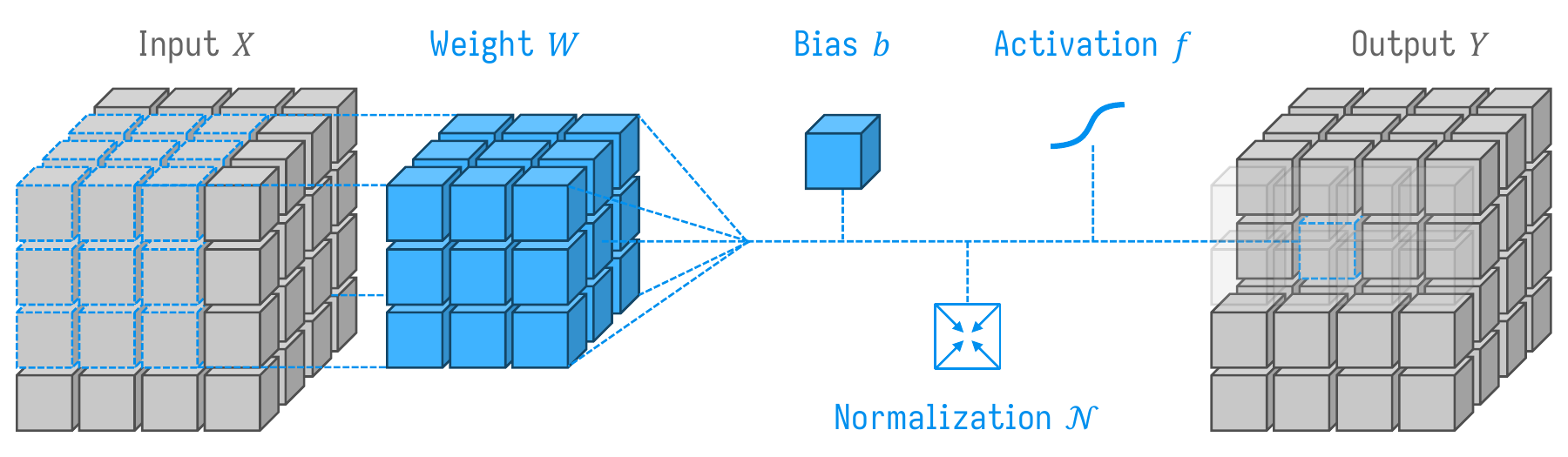}
  \caption{Schematic of a convolutional block used within the network. The process flow shows the input tensor $X$ being convolved with weights $W$. Subsequent operations include the addition of a bias term $b$, normalization $\mathcal{N}$, and a nonlinear activation function $f$ to produce the output tensor $Y$.}
  \label{fig:convolutional-layer}
\end{figure}

The fundamental building block is a convolutional layer followed by optional normalization and nonlinear activation, as detailed in \autoref{fig:convolutional-layer}. Each block applies a $K \times K \times K$ convolution, an optional bias term, optional normalization $\mathcal{N}$, and an activation function $f$. The output layer omits the activation so that predictions span the full dynamic range required by the \num{36} tensor components. All convolutions are implemented with PyTorch \cite{paszke2019,ansel2024} \texttt{Conv3d} using circular padding to enforce the periodic boundary conditions inherent to our data.

Formally, let $X \in \mathbb{R}^{C_{\text{in}} \times D \times D \times D}$ denote the input tensor, where $C_{\text{in}}$ is the number of input channels and $D$ the spatial resolution. The layer is parameterized by weights $W \in \mathbb{R}^{C_{\text{out}} \times C_{\text{in}} \times K \times K \times K}$ and biases $b \in \mathbb{R}^{C_{\text{out}}}$, where $C_{\text{out}}$ is the number of output channels and $K$ the (odd) kernel size with half-width $\kappa = (K-1)/2$. Adopting the deep-learning convention of cross-correlation (commonly termed convolution) and enforcing periodic boundaries via the modulo operator, the operation at spatial position $(x,y,z)$ for output channel $c_{\text{out}}$ reads:
\begin{equation}
  \text{Conv3D}(X)_{c_{\text{out}}, x, y, z} = b_{c_{\text{out}}} + \sum_{c_{\text{in}}=0}^{C_{\text{in}}-1} \sum_{i=0}^{K-1} \sum_{j=0}^{K-1} \sum_{k=0}^{K-1} W_{c_{\text{out}}, c_{\text{in}}, i, j, k} \cdot X_{c_{\text{in}},\, (x + i - \kappa) \bmod D,\, (y + j - \kappa) \bmod D,\, (z + k - \kappa) \bmod D}
  \label{eq:conv3d}
\end{equation}
The output tensor $Y \in \mathbb{R}^{C_{\text{out}} \times D \times D \times D}$ is then obtained by sequentially applying normalization and activation:
\begin{equation}
  Y = f \left( \mathcal{N} \left( \text{Conv3D}(X) \right) \right)
  \label{eq:layer-op}
\end{equation}

The activation functions explored during the hyperparameter tuning are:

\begin{equation}
  \mathrm{LeakyReLU}(x) = \max(0.01 \cdot x, x)
  \label{eq:leakyrelu}
\end{equation}

\begin{equation}
  \mathrm{Sigmoid}(x) = \frac{1}{1 + e^{-x}}
  \label{eq:sigmoid}
\end{equation}

\begin{equation}
  \mathrm{SiLU}(x) = \frac{x}{1 + e^{-x}}
  \label{eq:silu}
\end{equation}

\FloatBarrier

\begin{figure}[ht]
  \centering
  \includegraphics[width=1\textwidth]{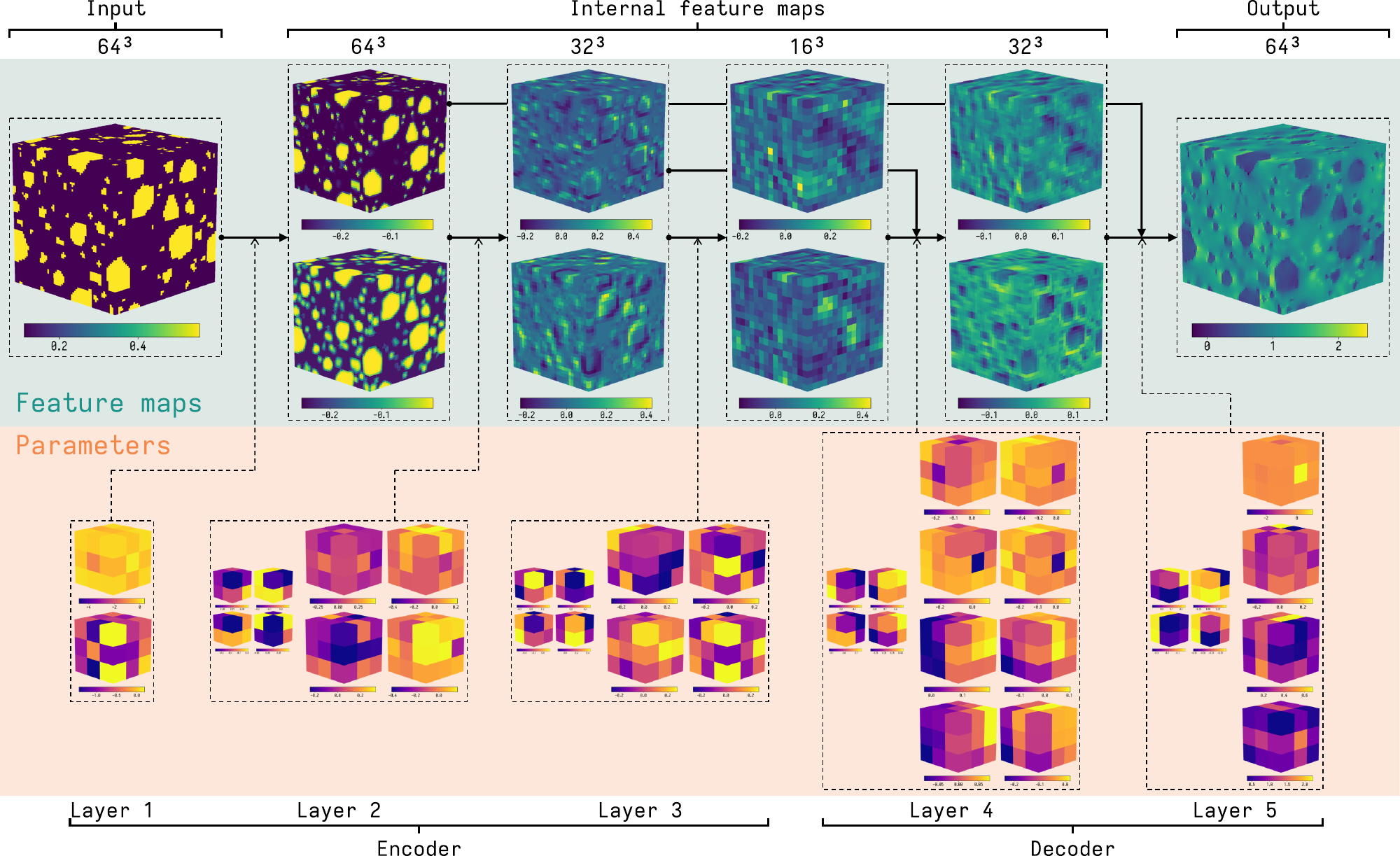}
  \caption{Internal structure of a representative U-Net instance for $64^3$ inputs. Top row: Feature maps through the U-Net. Bottom row: Learned convolutional kernels for layers $1$--$5$, illustrating the hierarchical extraction of features from fine-scale textures to coarse morphological patterns.}
  \label{fig:architecture-visualization}
\end{figure}

\autoref{fig:architecture-visualization} illustrates the architecture of a U-Net configured for $64^3$ inputs, predicting the $\mathbb{A}_{1111}$ component of the strain localization tensor. This low-dimensional configuration is used as an illustrative example. The top row shows feature-map evolution, where spatial resolution contracts from $64^3$ to $16^3$ in the encoder, before reversing in the decoder. The bottom row displays learned convolutional kernels at successive depths. Such hierarchical, multiscale feature extraction is essential for resolving both localized strain concentrations and long-range elastic interactions. The small $2^3$ kernels are used for downsampling and upsampling using standard and transposed convolutions, respectively, with stride 2. The larger $3^3$ kernels are used in the main convolution operations at constant resolution. Hyperparameter values for this illustrative example are listed in \autoref{tab:visualization-HPs}. While neural networks remain largely ``black boxes'', this visualization provides intuition on the input-output transformation and illustrates how some hyperparameters can influence performance, computational cost, and memory usage.

\FloatBarrier

\begin{table}[ht]
  \centering
  \caption{Hyperparameters of the neural network illustrated in \autoref{fig:architecture-visualization}.}
  \label{tab:visualization-HPs}
  \begin{tabular}{lc}
    \toprule
    \textbf{Hyperparameter} & \textbf{Value} \\
    \midrule
    Number of channels & $2$ \\
    Number of levels & $3$ \\
    Number of layers per level & $1$ \\
    Kernel size & $3$ \\
    Separated convolution & \texttt{False} \\
    Constant channels & \texttt{True} \\
    Bias & \texttt{False} \\
    Skip connections & \texttt{Direct} \\
    Normalization method & \texttt{None} \\
    Activation function & \texttt{SiLU} \\
    \bottomrule
  \end{tabular}
\end{table}
\FloatBarrier

In \autoref{sec:results}, these hyperparameters are optimized through random search, but first, the training process is described in the following section.

\subsection{Training}
\label{sec:training}

\subsubsection{Loss function}
\label{sec:loss}

The neural network is optimized following standard deep learning protocols. The training objective is to minimize a scalar loss function that quantifies the discrepancy between the model predictions and the reference data considered as ``ground truth''. Gradients of the loss with respect to the learnable parameters are computed via automatic differentiation, facilitating iterative updates through backpropagation.

For this study, we employ the Mean Absolute Error (MAE) as the objective function:
\begin{equation}
  \mathcal{L}_{\text{MAE}}(Y, \hat{Y}) = \frac{1}{B \cdot C \cdot D^3} \sum_{b,c,x,y,z} \left| Y(b, c, x, y, z) - \hat{Y}(b, c, x, y, z) \right|
  \label{eq:loss}
\end{equation}
where $B$ denotes the batch size, $C$ represents the number of output components (\num{36} channels), and $D$ is the spatial resolution. The indices $b$, $c$, $x$, $y$, and $z$ correspond to the batch, component, and spatial coordinates, respectively. Here, $Y$ represents the ground truth tensor field, and $\hat{Y}$ denotes the network prediction.

Because the six macroscopic loadings are unit basis vectors (see \autoref{sec:data-out}), each component of $\mathbb{A}$ is dimensionless and the absolute error in \autoref{eq:loss} is already relative to a unit macroscopic strain. The percentage errors reported throughout this work are therefore obtained by multiplying \autoref{eq:loss} by 100, i.e., they represent the mean absolute deviation between the predicted and reference localization tensors, expressed in percent. The reported percentage values are absolute deviations of the dimensionless localization tensor, expressed in percent of unit macroscopic strain, not relative errors.

The MAE formulation ensures a spatially uniform penalization of the absolute error, penalizing absolute deviations linearly across all voxels and tensor components.

This loss is minimized using one of the optimization algorithms described in the following section.

\subsubsection{Optimizer}
\label{sec:optimizer}

The optimization process aims to find the set of neural network parameters $\theta$ that minimizes the loss function $\mathcal{L}$ defined in \autoref{sec:loss}. The choice of optimizer significantly influences the convergence rate, the stability of the training process, and the final generalization error. In the context of predicting physical fields, where the loss landscape can be complex due to the high-dimensional output space (\num{36} components) and the heterogeneous nature of the input microstructures, selecting an appropriate optimization algorithm is critical. We evaluate three distinct optimizers ranging from classical stochastic methods to advanced techniques incorporating matrix orthogonalization.

Stochastic Gradient Descent (SGD) with momentum is a foundational optimization algorithm in deep learning \cite{polyak1964}. While standard SGD can suffer from slow convergence and oscillations in directions of high curvature, the addition of a momentum term accumulates a velocity vector in directions of persistent reduction in the loss. This helps the optimizer navigate through ravines in the loss landscape and dampen oscillations. It serves as a robust baseline for comparison, often providing good generalization properties despite slower convergence compared to adaptive methods. The update rule is detailed in \autoref{alg:sgd}, where $l_r$ is the learning rate and $\beta$ is the momentum coefficient.

\begin{algorithm}
  \caption{Stochastic gradient descent with momentum}
  \label{alg:sgd}
  \begin{algorithmic}
    \State \textbf{Hyperparameters:} $l_r, \beta$
    \State \textbf{Iteration:}
    \State $G_t = \nabla_\theta \mathcal{L}_t(\theta_{t-1})$
    \State $m_t = \beta m_{t-1} + G_t$
    \State $\theta_t = \theta_{t-1} - l_r m_t$
  \end{algorithmic}
\end{algorithm}
\FloatBarrier

AdamW \cite{loshchilov2019} is a variant of the Adam optimizer \cite{kingma2017} that implements decoupled weight decay regularization. Adam adapts the learning rate for each parameter based on the first and second moments of the gradients, allowing for efficient navigation of sparse gradients and noisy objectives. However, standard Adam couples L2 regularization with the gradient update, which can lead to suboptimal regularization behavior. AdamW decouples this, applying weight decay $\lambda$ directly to the weights rather than the gradient. This method is widely adopted in modern deep learning architectures due to its fast convergence and stability. The specific update steps, including bias correction for the moment estimates ($\hat{m}_t, \hat{v}_t$), are presented in \autoref{alg:adamw}.

\begin{algorithm}
  \caption{AdamW}
  \label{alg:adamw}
  \begin{algorithmic}
    \State \textbf{Hyperparameters:} $l_r, \beta_1, \beta_2, \lambda, \epsilon$
    \State \textbf{Iteration:}
    \State $G_t = \nabla_\theta \mathcal{L}_t(\theta_{t-1})$
    \State $m_t = \beta_1 m_{t-1} + (1 - \beta_1) G_t$
    \State $v_t = \beta_2 v_{t-1} + (1 - \beta_2) G_t^2$
    \State $\hat{m}_t = \frac{m_t}{1 - \beta_1^t}, \quad \hat{v}_t = \frac{v_t}{1 - \beta_2^t}$
    \State $\theta_t = \theta_{t-1} - l_r \left( \frac{\hat{m}_t}{\sqrt{\hat{v}_t} + \epsilon} + \lambda \theta_{t-1} \right)$
  \end{algorithmic}
\end{algorithm}
\FloatBarrier

The Muon optimizer \cite{jordan2024} represents a more recent class of optimization algorithms that incorporate second-order information or matrix structural constraints to improve stability. As illustrated in \autoref{alg:muon}, Muon computes the momentum buffer similarly to SGD but applies a Newton-Schulz (NS) iteration, denoted as $\text{NS}(\cdot)$, to the momentum matrix $m_t$ before the parameter update. This orthogonalization step aims to stabilize the update direction, potentially preventing the exploding or vanishing gradient problems that can occur in deep networks with complex physical constraints. By enforcing orthogonality on the update matrix, Muon seeks to maintain a well-conditioned optimization trajectory, which may be particularly beneficial for the high-dimensional tensor field predictions targeted in this study. In practice, we employ a hybrid optimization strategy to accommodate the structural requirements of the Muon algorithm. Muon is exclusively applied to the convolutional kernels of the intermediate hidden layers, whereas AdamW is utilized for all other parameters, including the input and output layer weights and any optional bias terms. Since the convolutional kernels defined in \autoref{eq:conv3d} are 5D tensors ($C_{\text{out}} \times C_{\text{in}} \times K \times K \times K$), they must be reshaped to facilitate the matrix orthogonalization inherent to the Newton-Schulz iteration. Specifically, the last four dimensions are flattened to convert the update tensors into matrices before orthogonalization, and folded back into their original shape before the parameter update. This procedure ensures compatibility with the convolutional operation while benefiting from the stability of orthogonal updates.

\begin{algorithm}
  \caption{Muon}
  \label{alg:muon}
  \begin{algorithmic}
    \State \textbf{Hyperparameters:} $l_r, \beta$
    \State \textbf{Iteration:}
    \State $G_t = \nabla_\theta \mathcal{L}_t(\theta_{t-1})$
    \State $m_t = \beta m_{t-1} + G_t$
    \State $O_t = \text{NS}(m_t)$
    \State $\theta_t = \theta_{t-1} - l_r O_t$
  \end{algorithmic}
\end{algorithm}
\FloatBarrier

\section{Results}
\label{sec:results}

\subsection{Hyperparameter tuning}
\label{sec:tuning}

To achieve an optimal trade-off between predictive accuracy and inference latency, a systematic hyperparameter optimization strategy was employed. The optimal configuration was identified through three sequential search stages, designed to progressively refine the search space. The initial search encompassed a broad parameter space with approximately \num{700} million possible combinations. Based on the validation MAE obtained from this coarse search, the parameter ranges were subsequently narrowed by excluding underperforming regions and fixing insensitive hyperparameters to values that minimize inference cost. Each of the three search stages comprised \num{1370} trials, where each trial involved training a model from scratch with random hyperparameter values sampled within the ranges defined in \autoref{tab:HP-configs}. Search 1 was performed only for the baseline configuration, whereas Search 2 and Search 3 were performed independently for the baseline and the three scaling configurations, i.e., $4 \times 1370$ trials per stage. To maintain computational feasibility, each trial was subject to a time-constrained training budget of \SI{300}{\second} (Search 1 and Search 2) and \SI{600}{\second} (Search 3) on a single Nvidia H100 GPU. Configurations deemed infeasible due to algorithmic constraints (e.g., convolution folding resulting from large kernels combined with excessive depth) or memory limitations (out-of-memory errors) were pruned from the search space.

\begin{table}[ht]
  \centering
  \caption{Evolution of the hyperparameter search space across three sequential refinement stages, transitioning from coarse exploration (Search 1) to constrained tuning (Search 2--3). The ranges in each column apply to each configuration. Search 1 covers the baseline configuration only; Search 2 and Search 3 cover the baseline and the three scaling configurations.}
  \label{tab:HP-configs}
  {\setlength{\tabcolsep}{1pt}
    \begin{tabular}{lccc}
      \toprule
      \textbf{} & \textbf{Search 1} & \textbf{Search 2} & \textbf{Search 3} \\
      \midrule
      \multicolumn{4}{l}{\textbf{Architecture hyperparameters}} \\
      \midrule
      \multirow{2}{*}{Number of channels} & $\left[2, 4, 8, 12, 16, 24,\right.$ & $\left[2, 4, 8, 12, 16, 24,\right.$ & $\left[8, 12, 16, 24, 32,\right.$ \\
      & $\left.32, 48, 64, 96, 128\right]$ & $\left.32, 48, 64, 96, 128\right]$ & $\left.32, 48, 64, 96, 128\right]$ \\
      Number of levels & $\left[1, 2, 3, 4, 5, 6, 7, 8\right]$ & $\left[3, 4, 5, 6, 7, 8\right]$ & $\left[3, 4, 5, 6, 7, 8\right]$ \\
      Number of layers per level & $\left[1, 2, 3, 4\right]$ & $1$ & $1$ \\
      Kernel size & $\left[2, 3, 4, 5, 6, 7\right]$ & $3$ & $3$ \\
      Separated convolution & $\left[\texttt{False}, \texttt{True}\right]$ & $\texttt{False}$ & $\texttt{False}$ \\
      Constant channels & $\left[\texttt{False}, \texttt{True}\right]$ & $\texttt{True}$ & $\texttt{True}$ \\
      Bias & $\left[\texttt{False}, \texttt{True}\right]$ & $\texttt{False}$ & $\texttt{False}$ \\
      Skip connections & $\left[\texttt{off}, \texttt{direct}, \texttt{trainable}\right]$ & $\texttt{direct}$ & $\texttt{direct}$ \\
      Normalization method & $\left[\texttt{off}, \texttt{instance}, \texttt{batch}\right]$ & $\left[\texttt{off}, \texttt{instance}, \texttt{batch}\right]$ & $\texttt{off}$ \\
      Activation function & $\left[\texttt{LeakyReLU}, \texttt{Sigmoid}, \texttt{SiLU}\right]$ & $\left[\texttt{LeakyReLU}, \texttt{Sigmoid}, \texttt{SiLU}\right]$ & $\texttt{SiLU}$ \\
      \midrule
      \multicolumn{4}{l}{\textbf{Training hyperparameters}} \\
      \midrule
      Optimizer & $\left[\texttt{SGD}, \texttt{AdamW}, \texttt{Muon}\right]$ & $\texttt{Muon}$ & $\texttt{Muon}$ \\
      Batch size & $\left[1, 2, 4, 8, 16\right]$ & $1$ & $1$ \\
      Learning rate & $\left[10^{-6}, 10^{0}\right]$ & $\left[10^{-6}, 10^{0}\right]$ & $\left[10^{-5}, 10^{-1}\right]$ \\
      \multirow{2}{*}{Momentum} & $\left[0.5, 0.7, 0.9, 0.95, 0.99,\right.$ & $\left[0.5, 0.7, 0.9, 0.95, 0.99,\right.$ & $\left[0.9, 0.95, 0.99,\right.$ \\
      & $\left.0.995, 0.999, 0.9999\right]$ & $\left.0.995, 0.999, 0.9999\right]$ & $\left.0.995, 0.999\right]$ \\
      \midrule
      Training budget & \SI{300}{\second} & \SI{300}{\second} & \SI{600}{\second} \\
      \bottomrule
  \end{tabular}}
\end{table}
\FloatBarrier

\begin{figure}[ht]
  \centering
  \includegraphics[width=1\textwidth]{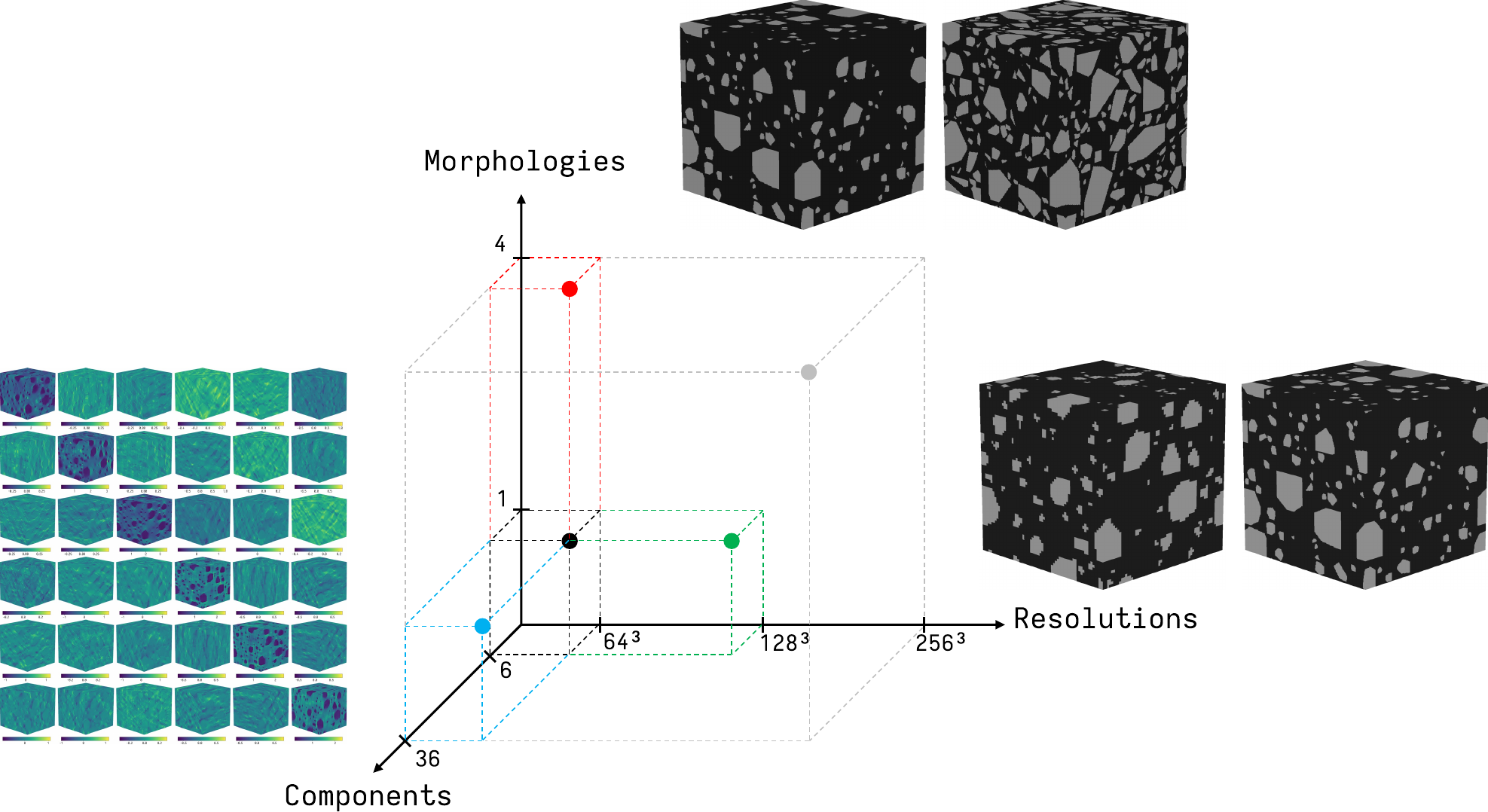}
  \caption{Schematic of the progressive scaling study mapping the trajectory from low-complexity baseline to high-complexity target across spatial resolution, output dimensionality, and morphological diversity. The black dot represents the baseline configuration ($64^3$, $6$ components, $1$ morphology), while the gray dot indicates the target configuration ($256^3$, $36$ components, $4$ morphologies). The colored dots (blue, green, red) represent intermediate configurations used to assess scaling behavior.}
  \label{fig:axes}
\end{figure}

To mitigate the computational cost associated with optimizing hyperparameters for the full-scale problem, a progressive scaling study was conducted. The optimization process began at a low-complexity baseline: a resolution of $64^3$, predicting only the first \num{6} components of the localization tensor, and utilizing a single type of microstructure morphology (denoted as the baseline configuration in \autoref{fig:axes}). Subsequently, the complexity was incrementally increased along three independent axes: spatial resolution, output dimensionality (number of tensor components), and morphological diversity. This approach, illustrated by the intermediate configurations in \autoref{fig:axes}, allowed for the evaluation of scaling laws to determine which optimal hyperparameters remain invariant and which require adjustment as problem complexity increases. These scaling rules were then applied to define the final training configuration for the high-complexity target scenario (resolution $256^3$, \num{36} components, and varied morphologies), represented by the target configuration in \autoref{fig:axes}.

The hyperparameter tuning study was performed using Optuna \cite{akiba2019} and MLflow \cite{chen2020}.

\FloatBarrier

\begin{figure}[ht]
  \centering
  \begin{subfigure}[b]{0.49\textwidth}
    \caption{}
    \label{fig:r64-c6-mat1-600s-accuracy-num_channels}
    \includegraphics[width=\textwidth]{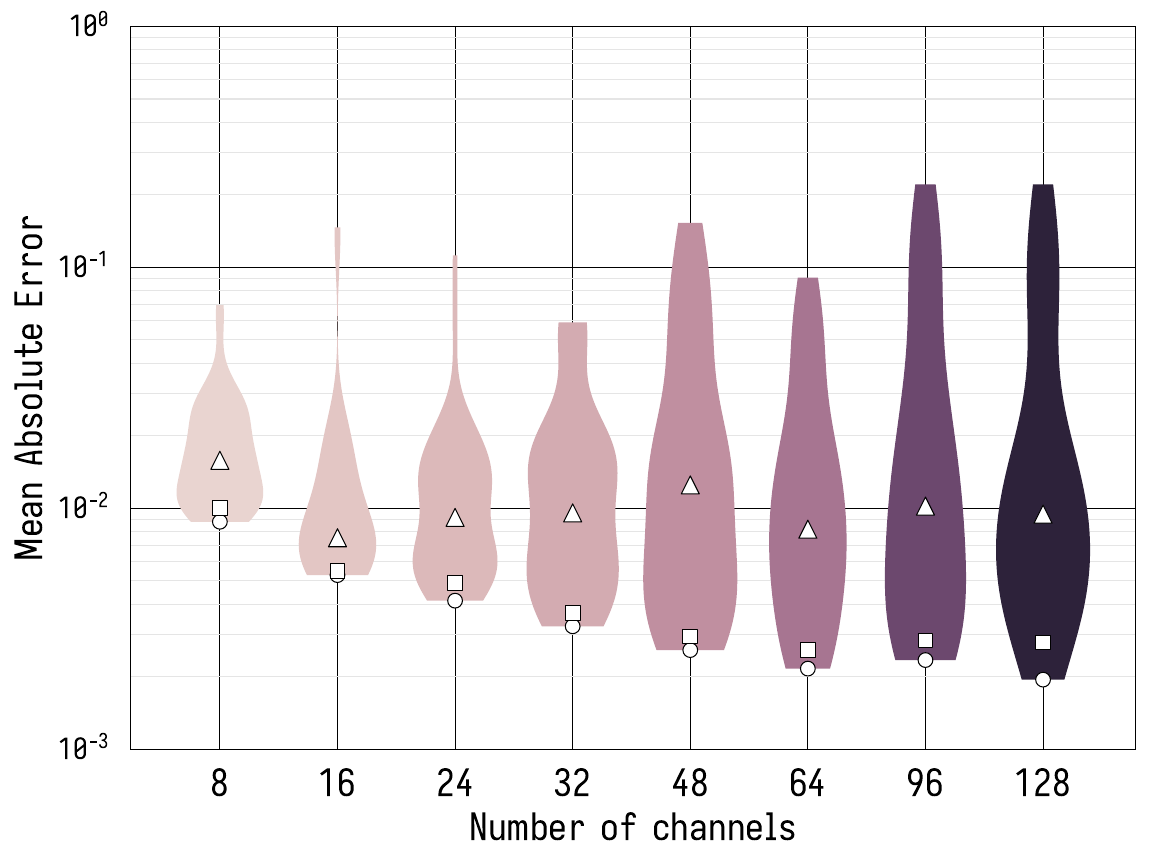}
  \end{subfigure}
  \hfill
  \begin{subfigure}[b]{0.49\textwidth}
    \caption{}
    \label{fig:r64-c6-mat1-600s-speed-num_channels}
    \includegraphics[width=\textwidth]{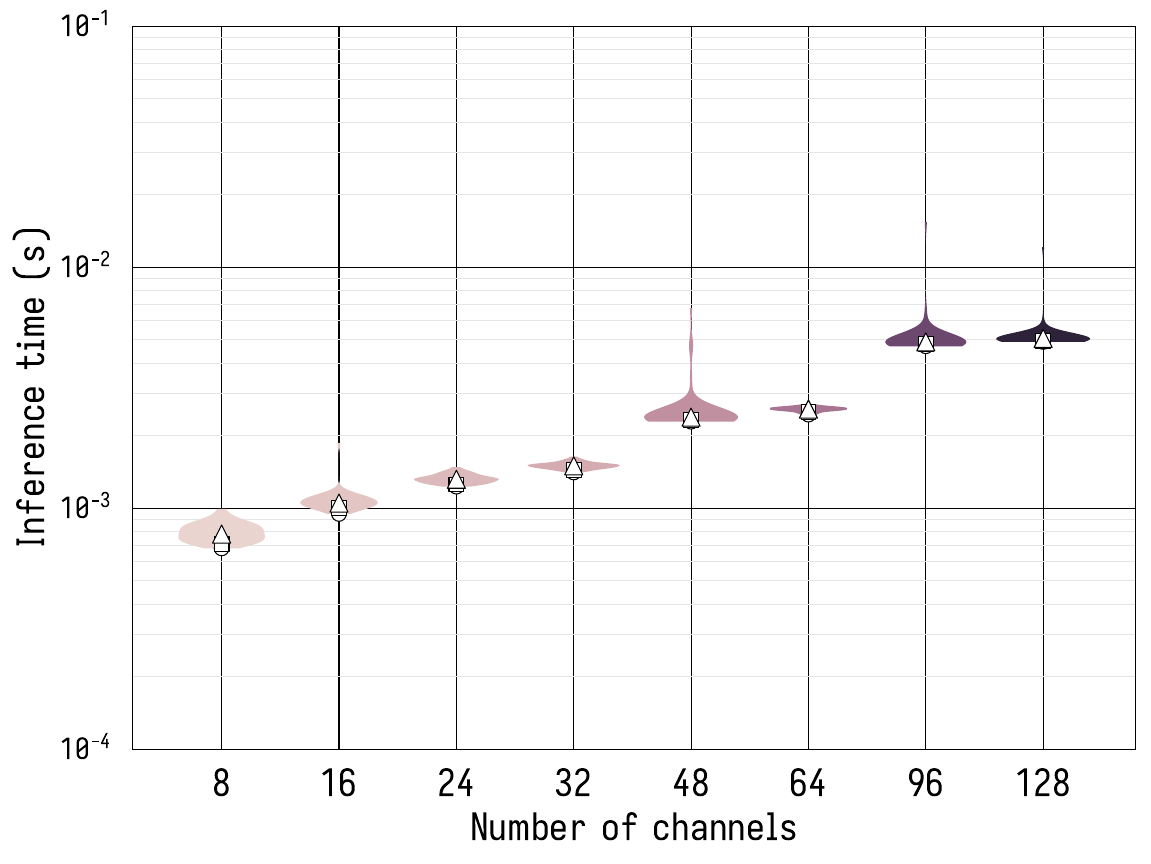}
  \end{subfigure}
  \hfill
  \begin{subfigure}[b]{0.49\textwidth}
    \caption{}
    \label{fig:r64-c6-mat1-600s-accuracy-num_levels}
    \includegraphics[width=\textwidth]{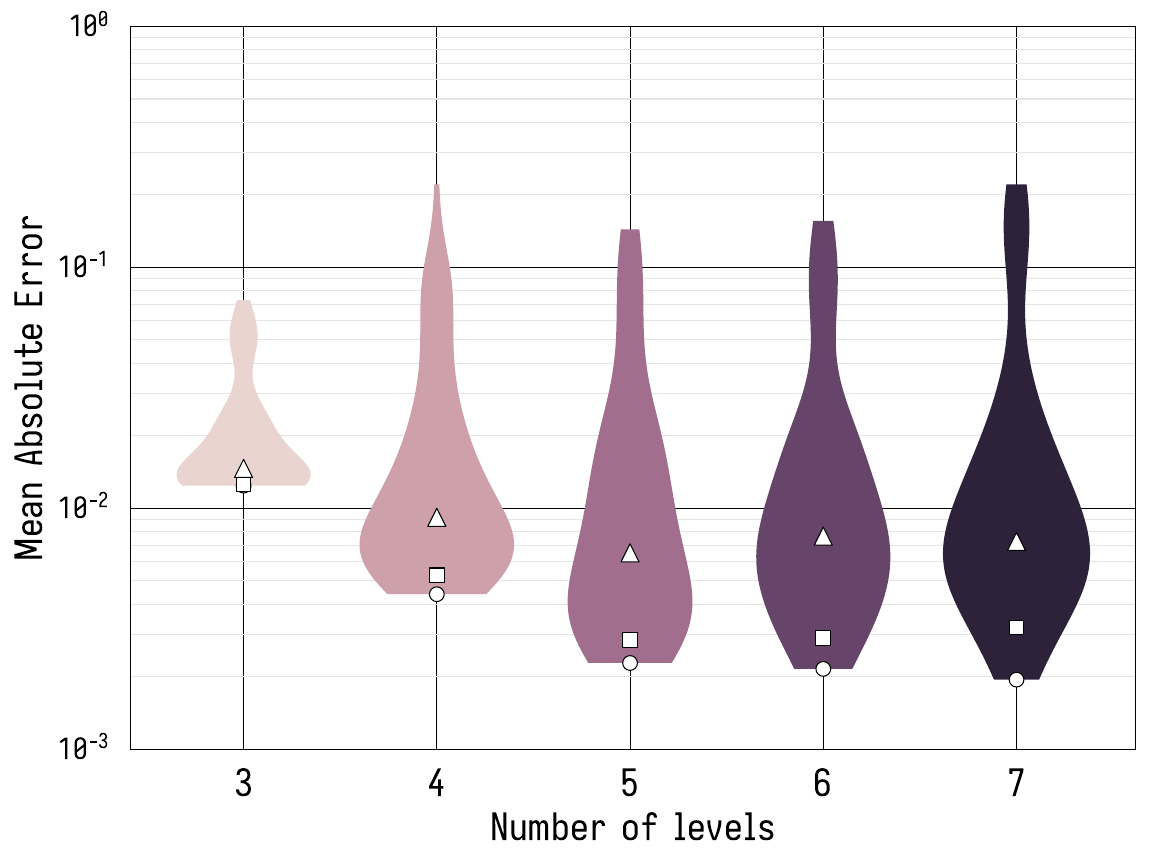}
  \end{subfigure}
  \hfill
  \begin{subfigure}[b]{0.49\textwidth}
    \caption{}
    \label{fig:r64-c6-mat1-600s-speed-num_levels}
    \includegraphics[width=\textwidth]{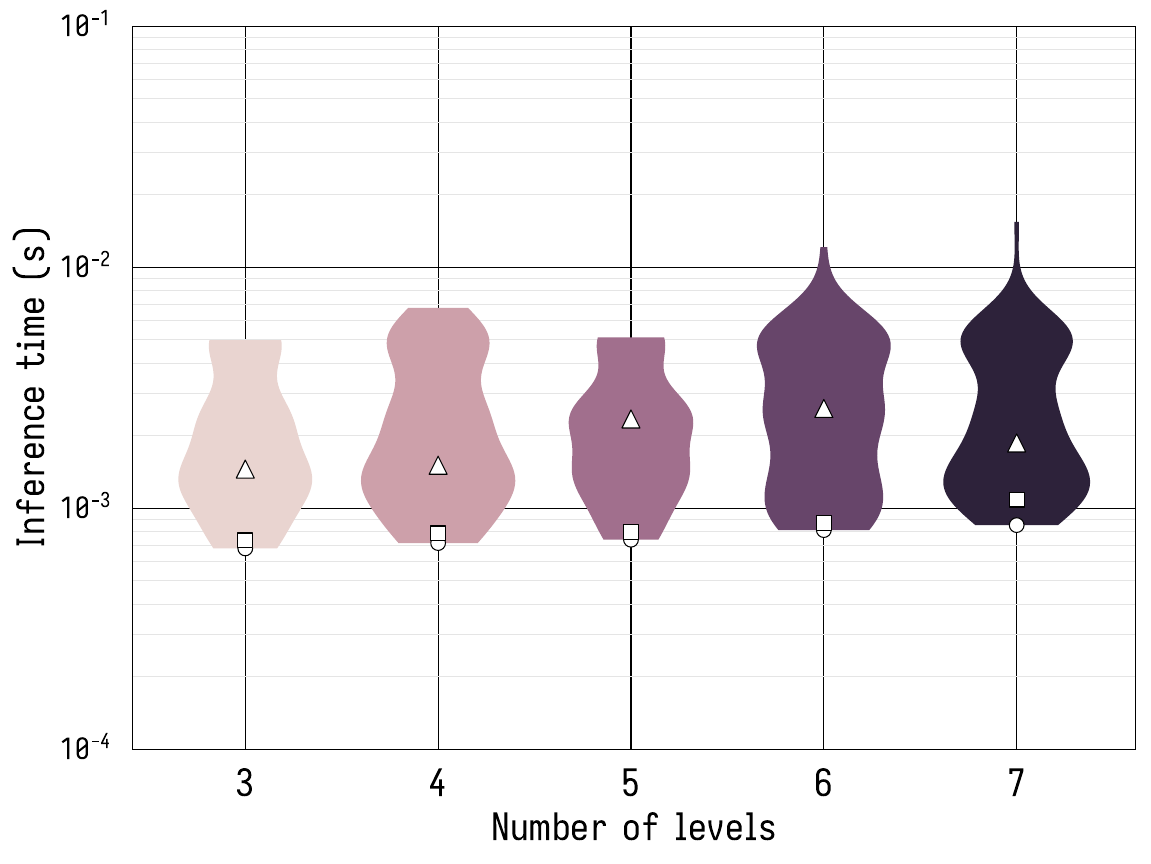}
  \end{subfigure}
  \hfill
  \begin{subfigure}[b]{0.49\textwidth}
    \caption{}
    \label{fig:r64-c6-mat1-600s-accuracy-learning_rate_exponent}
    \includegraphics[width=\textwidth]{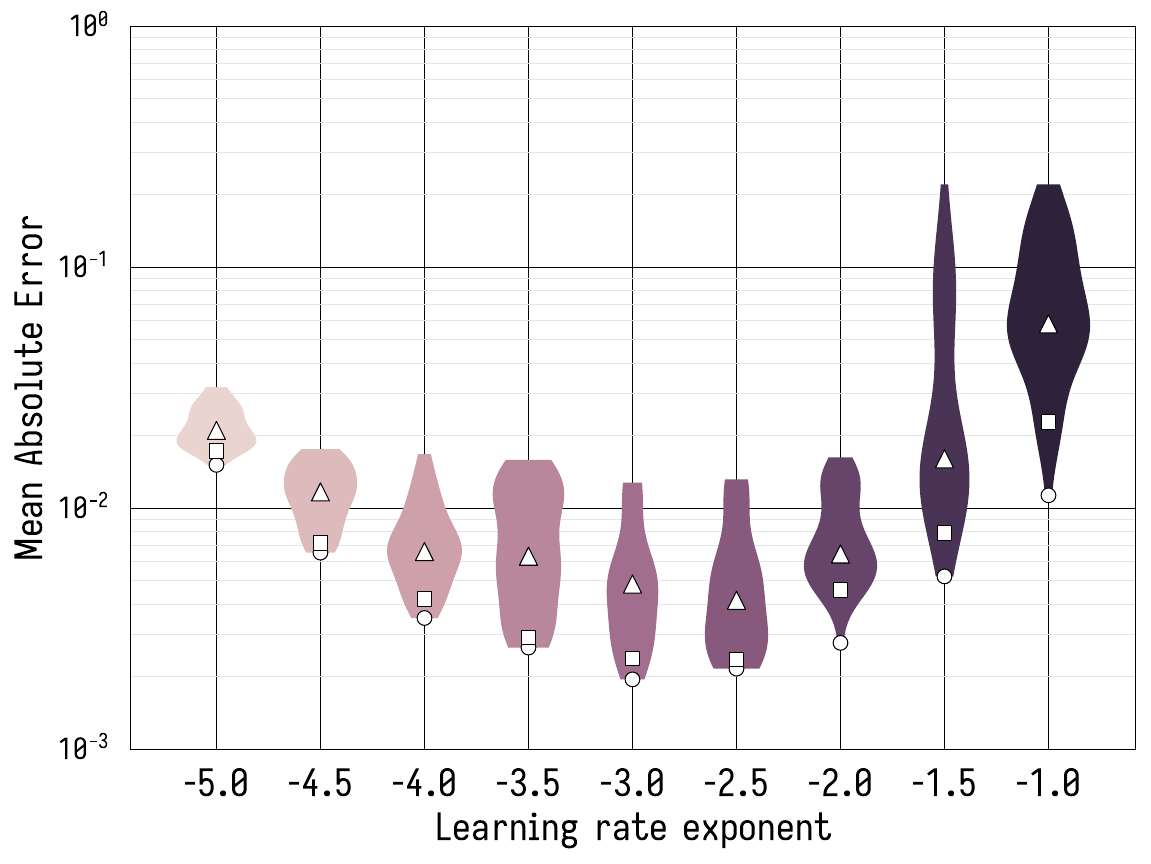}
  \end{subfigure}
  \hfill
  \begin{subfigure}[b]{0.49\textwidth}
    \caption{}
    \label{fig:r64-c6-mat1-600s-speed-learning_rate_exponent}
    \includegraphics[width=\textwidth]{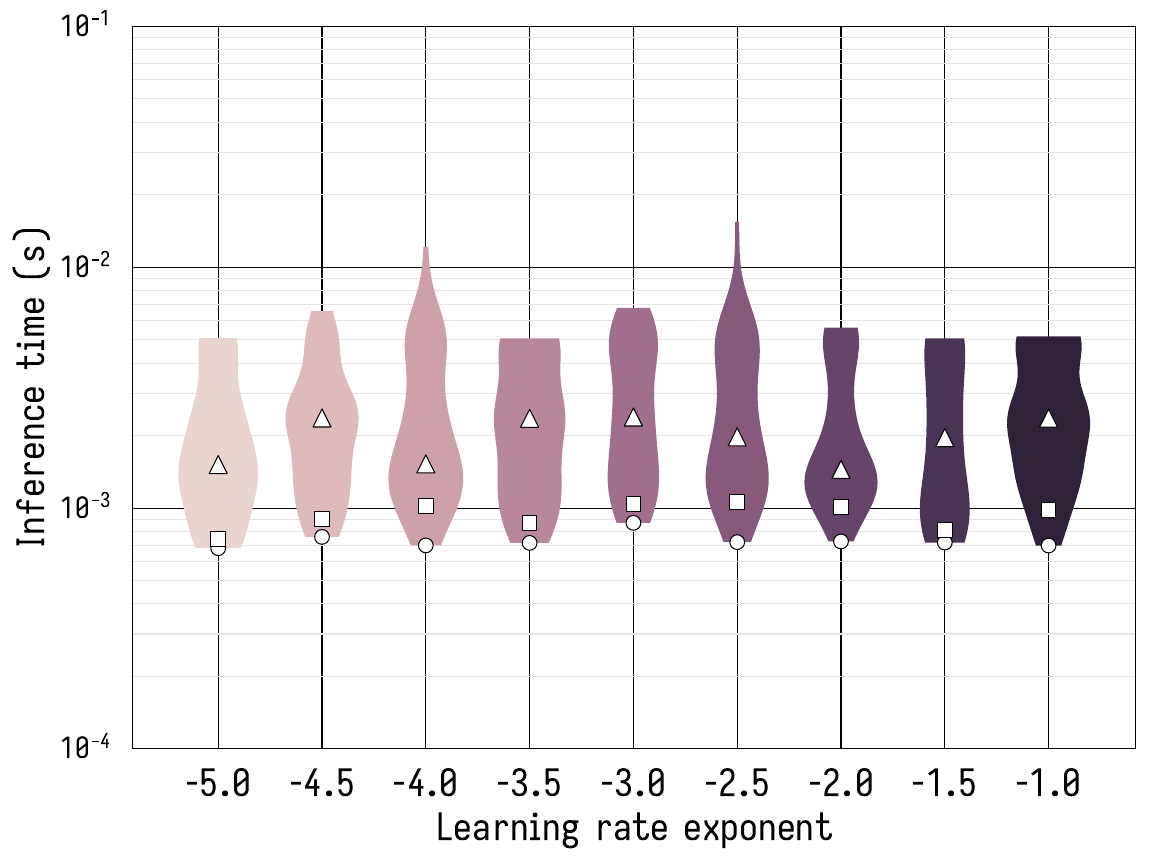}
  \end{subfigure}
  \caption{Hyperparameter tuning results for Search 3 of the baseline configuration ($64^3$ resolution, 6 output components, 1 morphology). (a, c, e) show the Mean Absolute Error (MAE), while (b, d, f) show the inference time per sample. The plots illustrate the sensitivity of model performance to the number of channels, number of levels, and learning rate exponent. The violin plots are obtained using Seaborn \cite{waskom2021}.}
  \label{fig:r64-c6-mat1-600s}
\end{figure}

The results of the hyperparameter optimization for the baseline configuration ($64^3$ resolution, 6 output components, and single morphology, Search 3) are presented in \autoref{fig:r64-c6-mat1-600s}. Due to the extensive search space, we focus on three key hyperparameters for this baseline configuration to illustrate the general trends obtained; the complete set of graphs for all searches is provided in the supplementary material. In the figures, triangles indicate the 50th percentile (median), squares represent the 10th percentile, and circles denote the minimum values observed across trials. These minimum values are particularly significant as they represent the best performance achievable by a model configuration for each specific hyperparameter value. The analysis evaluates the trade-off between predictive accuracy, measured by the Mean Absolute Error (MAE), and computational efficiency, measured by the inference time per sample.

\autoref{fig:r64-c6-mat1-600s} panels (a) and (b) illustrate the impact of the number of channels. A clear decrease in MAE is observed up to \num{64} channels, beyond which only marginal gains are achieved. Regarding computational cost, inference time increases monotonically with the number of channels. Consequently, the number of channels serves as an effective hyperparameter to balance accuracy against speed.

The influence of network depth, represented by the number of levels, is shown in \autoref{fig:r64-c6-mat1-600s}(c, d). A clear trend emerges where increasing the number of levels from \num{3} to \num{5} consistently reduces the MAE, with marginal gains observed above \num{5} levels. Notably, this improvement in accuracy is achieved with minimal impact on inference time, which remains relatively stable across different depths. At this resolution, \num{5} levels represent the ideal balance, indicating that increasing the depth of the U-Net is an efficient strategy for enhancing performance without significantly compromising speed.

Finally, the sensitivity to the learning rate is depicted in \autoref{fig:r64-c6-mat1-600s}(e, f). The MAE exhibits a strong dependence on the learning rate exponent. Values that are too small ($l_r \leq 10^{-4.5}$) result in slower convergence and higher error, while values that are too large ($l_r \geq 10^{-1.5}$) lead to instability and a drastic increase in error. The optimal range is found between learning rates $10^{-4.0}$ and $10^{-2.0}$, where the model achieves the best compromise between accuracy and stability. Inference time remains largely unaffected by the learning rate, as expected, since the learning rate governs the training dynamics rather than the computational graph used during inference.

\FloatBarrier

The optimal hyperparameters presented in \autoref{tab:HP-accuracy-optimal} and \autoref{tab:HP-speed-optimal} were selected from the full set of trials in Search 3 across each configuration. The selection followed two distinct criteria to address different application needs:
\begin{enumerate}
  \item \textbf{Accuracy-optimal:} the configuration yielding the minimum validation MAE, regardless of inference time.
  \item \textbf{Speed-optimal:} the configuration yielding the minimum inference time, subject to the constraint that the validation MAE remains lower than \SI{1}{\percent}.
\end{enumerate}
This dual-criterion approach allows users to tailor the surrogate model based on whether the priority is maximum fidelity or computational throughput.

Analysis of the accuracy-optimal configurations in \autoref{tab:HP-accuracy-optimal} reveals several trends regarding the scaling of network capacity with problem complexity. For the baseline configuration (\texttt{r64-c6-m1}), the optimal model utilizes \num{128} channels and \num{7} levels, achieving a remarkably low error of \SI{0.19}{\percent}. When the resolution is increased to $128^3$ (\texttt{r128-c6-m1}), the optimal number of channels decreases to \num{64}, while the depth remains at \num{7} levels. Investigation of the \num{128}-channel trials reveals performance metrics very close to those of the 64-channel trials, corroborating the finding discussed in \autoref{sec:tuning} that accuracy gains are marginal above 64 channels. Increasing the output dimensionality from \num{6} to \num{36} components (\texttt{r64-c36-m1}) requires maintaining the high channel count (\num{128}) but results in a slightly higher error (\SI{0.38}{\percent}), indicating the increased difficulty of mapping multiple tensor components simultaneously. Similarly, increasing morphological diversity from \num{1} to \num{4} configurations (\texttt{r64-c6-m4}) reduces the optimal depth to \num{5} levels. Investigation shows that \num{5} or more levels yield similar results, consistent with the earlier observation that marginal gains are achieved above \num{5} levels.

\begin{table}[ht]
  \centering
  \caption{Accuracy-optimal hyperparameters for different configurations.}
  \label{tab:HP-accuracy-optimal}
  \begin{tabular}{lcccc}
    \toprule
    & \texttt{r64-c6-m1} & \texttt{r128-c6-m1} & \texttt{r64-c36-m1} & \texttt{r64-c6-m4} \\

    \midrule
    \multicolumn{5}{l}{\textbf{Architecture hyperparameters}} \\
    \midrule

    Number of channels & $128$ & $64$ & $128$ & $128$ \\
    Number of levels & $7$ & $7$ & $7$ & $5$ \\

    \midrule
    \multicolumn{5}{l}{\textbf{Training hyperparameters}} \\
    \midrule

    Learning rate & $10^{-3}$ & $10^{-3}$ & $3 \cdot 10^{-3}$ & $10^{-3}$ \\
    Momentum & $0.999$ & $0.995$ & $0.99$ & $0.999$ \\

    \midrule
    \multicolumn{5}{l}{\textbf{Metrics}} \\
    \midrule

    Mean Absolute Error & \SI{0.19}{\percent} & \SI{0.35}{\percent} & \SI{0.38}{\percent} & \SI{0.38}{\percent} \\
    Inference time per sample & \SI{5.142}{\milli\second} & \SI{19.355}{\milli\second} & \SI{7.070}{\milli\second} & \SI{5.037}{\milli\second} \\

    \bottomrule
  \end{tabular}
\end{table}
\FloatBarrier

The speed-optimal configurations in \autoref{tab:HP-speed-optimal} demonstrate a significant reduction in model capacity to meet the inference time constraints. The number of channels drops drastically to between \num{8} and \num{24}, and the number of levels is reduced to between \num{4} and \num{6}. Despite this reduction, the error remains bounded just below the \SI{1}{\percent} threshold (ranging from \SI{0.93}{\percent} to \SI{0.98}{\percent}). The inference time improvements are substantial; for the baseline configuration, the speed-optimized model is approximately \num{6.6} times faster than the accuracy-optimized model (\SI{0.779}{\milli\second} vs \SI{5.142}{\milli\second}), highlighting the efficiency gains achievable when strict accuracy requirements are relaxed slightly.

In addition to the primary architectural dimensions of width and depth, we systematically evaluate several secondary hyperparameters to identify configurations that balance performance with computational efficiency. Regarding convolutional geometry, small odd-sized kernels (e.g., $K=3$) consistently outperform larger or even-sized kernels, whereas separated convolutional layers yield suboptimal performance. Structural choices also influenced efficiency: maintaining a constant channel count across levels yielded accuracy comparable to doubling channels upon downsampling but with a reduced memory footprint. Bias terms were excluded as they provided no measurable improvement, and direct skip connections proved superior to both absent and trainable skip connection topologies. Normalization layers (batch or instance) offered negligible benefits over no normalization, favoring the latter for computational simplicity. Among activation functions, SiLU yielded marginally lower errors than LeakyReLU, with Sigmoid performing worst. Regarding optimization, the Muon optimizer demonstrated superior performance relative to AdamW and SGD with momentum, characterized by enhanced stability in learning curves throughout training iterations and reduced final validation errors. A monotonic increase in validation error was observed with increasing batch sizes. Finally, optimal learning rate and momentum values were identified within specific ranges and remained invariant despite increases in problem complexity.

In conclusion, this analysis demonstrates that hyperparameters can be effectively optimized for either accuracy or speed based on specific user requirements. For a fixed objective, most optimal hyperparameters remain stable when scaling problem complexity (resolution, components, morphologies). Only two parameters require significant adjustment:
\begin{enumerate}
  \item The \textbf{number of channels} should be increased or decreased to control the accuracy versus speed tradeoff.
  \item The \textbf{number of levels} should increase with the resolution such that the minimum resolution of the feature maps (at the end of the encoder) reaches $4^3$.
\end{enumerate}

\begin{table}[ht]
  \centering
  \caption{Speed-optimal hyperparameters for different configurations.}
  \label{tab:HP-speed-optimal}
  \begin{tabular}{lcccc}
    \toprule
    & \texttt{r64-c6-m1} & \texttt{r128-c6-m1} & \texttt{r64-c36-m1} & \texttt{r64-c6-m4} \\

    \midrule
    \multicolumn{5}{l}{\textbf{Architecture hyperparameters}} \\
    \midrule

    Number of channels & $8$ & $8$ & $24$ & $16$ \\
    Number of levels & $5$ & $6$ & $4$ & $5$ \\

    \midrule
    \multicolumn{5}{l}{\textbf{Training hyperparameters}} \\
    \midrule

    Learning rate & $3 \cdot 10^{-4}$ & $10^{-3}$ & $3 \cdot 10^{-3}$ & $10^{-3}$ \\
    Momentum & $0.95$ & $0.99$ & $0.999$ & $0.995$ \\

    \midrule
    \multicolumn{5}{l}{\textbf{Metrics}} \\
    \midrule

    Mean Absolute Error & \SI{0.95}{\percent} & \SI{0.97}{\percent} & \SI{0.98}{\percent} & \SI{0.93}{\percent} \\
    Inference time per sample & \SI{0.779}{\milli\second} & \SI{5.385}{\milli\second} & \SI{3.033}{\milli\second} & \SI{1.033}{\milli\second} \\

    \bottomrule
  \end{tabular}
\end{table}
\FloatBarrier

Applying these guidelines, we now train a single model on the full target problem.

\subsection{Final training run}
\label{sec:res-final}

The final training run targets the high-complexity configuration identified in \autoref{fig:axes}, namely a spatial resolution of $256^3$ voxels, the full $36$-component localization tensor output, and the full morphological diversity comprising the four aggregate volume fraction and aspect ratio combinations described in \autoref{sec:data-in}.

Based on the hyperparameter tuning analysis presented in \autoref{sec:tuning}, a final configuration was selected to balance the competing objectives of predictive accuracy and inference speed. The chosen compromise utilizes \num{48} channels to balance accuracy and speed, and \num{7} levels leading to a minimum feature map resolution of $4^3$. The complete set of final hyperparameters is detailed in \autoref{tab:HP-final}.

\begin{table}[ht]
  \centering
  \caption{Final hyperparameters selected for the target configuration: $256^3$ resolution, 36 localization-tensor components, four morphologies.}
  \label{tab:HP-final}
  {\setlength{\tabcolsep}{1pt}
    \begin{tabular}{lc}
      \toprule
      \textbf{Hyperparameter} & \textbf{Value} \\
      \midrule
      Number of channels & $48$ \\
      Number of levels & $7$ \\
      Number of layers per level & $1$ \\
      Kernel size & $3$ \\
      Separated convolution & $\texttt{False}$ \\
      Constant channels & $\texttt{True}$ \\
      Bias & $\texttt{False}$ \\
      Skip connections & $\texttt{direct}$ \\
      Normalization method & $\texttt{off}$ \\
      Activation function & $\texttt{SiLU}$ \\
      \midrule
      Optimizer & $\texttt{Muon}$ \\
      Batch size & $1$ \\
      Learning rate & $10^{-3}$ \\
      Momentum & $0.999$ \\
      \bottomrule
  \end{tabular}}
\end{table}
\FloatBarrier

The selected architecture contains approximately \num{1.4} million trainable parameters. Peak GPU memory consumption during training reached approximately \SI{90}{\giga\byte}, which remains within the capacity of a single Nvidia H100 GPU but illustrates the substantial memory requirements of high-resolution 3D image-to-image regression.

\begin{figure}[ht]
  \centering
  \includegraphics[width=0.7\textwidth]{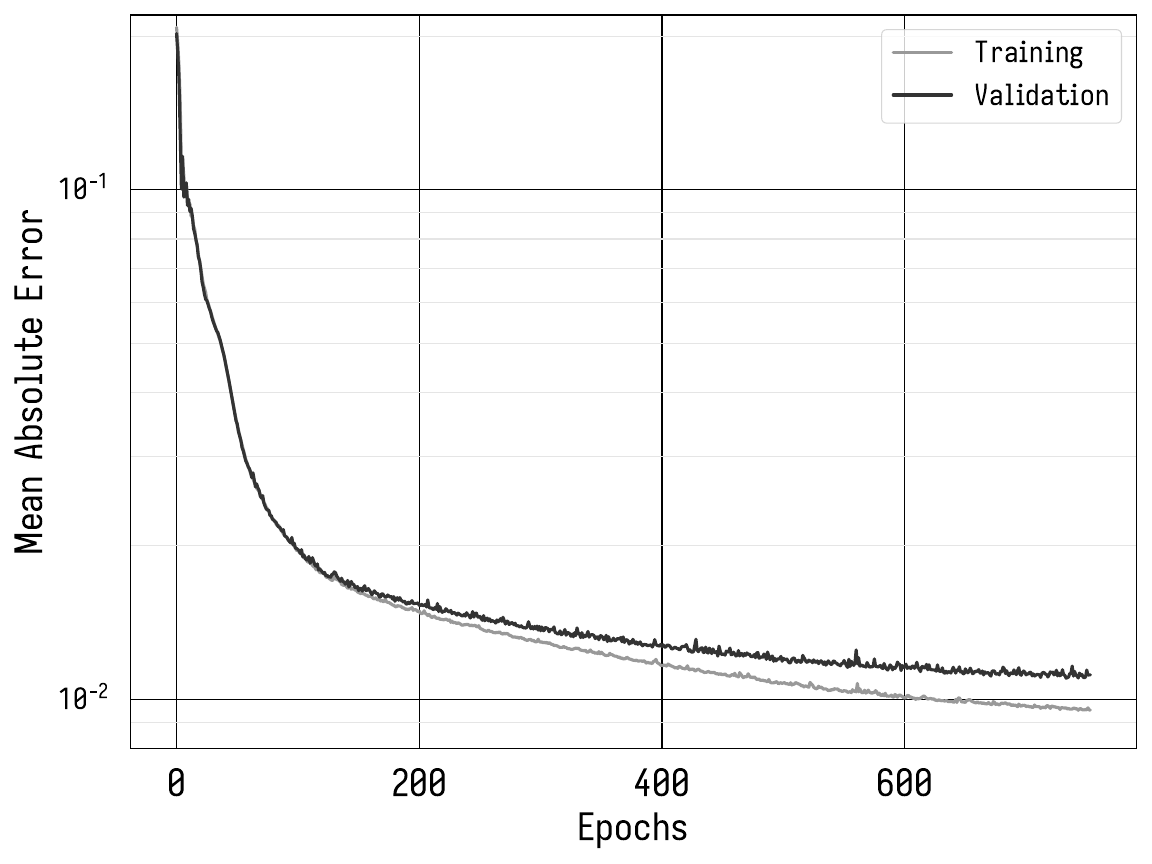}
  \caption{Learning curves showing the evolution of training and validation errors over epochs.}
  \label{fig:curves}
\end{figure}

The training dynamics for this configuration are illustrated in \autoref{fig:curves}. The model demonstrates rapid initial convergence, with the Mean Absolute Error decreasing below \SI{2}{\percent} within approximately \num{100} epochs, corresponding to a training time of roughly \SI{35}{\minute}. Following this initial phase, the training and validation errors begin to separate slightly. The training process was terminated once the validation error had stagnated, after a total training time of \SI{240}{\minute}, reaching a final training error of \SI{0.95}{\percent}, a validation error of \SI{1.1180}{\percent}, and a test error of \SI{1.1184}{\percent}. The test error was computed once on the held-out samples, after training and model selection were complete; its agreement with the validation error to within $0.0004$ percentage points indicates that the validation-based stopping and hyperparameter choices did not materially bias the reported performance.

\FloatBarrier

\begin{figure}[ht]
  \centering
  \begin{subfigure}[b]{0.32\textwidth}
    \includegraphics[width=\textwidth]{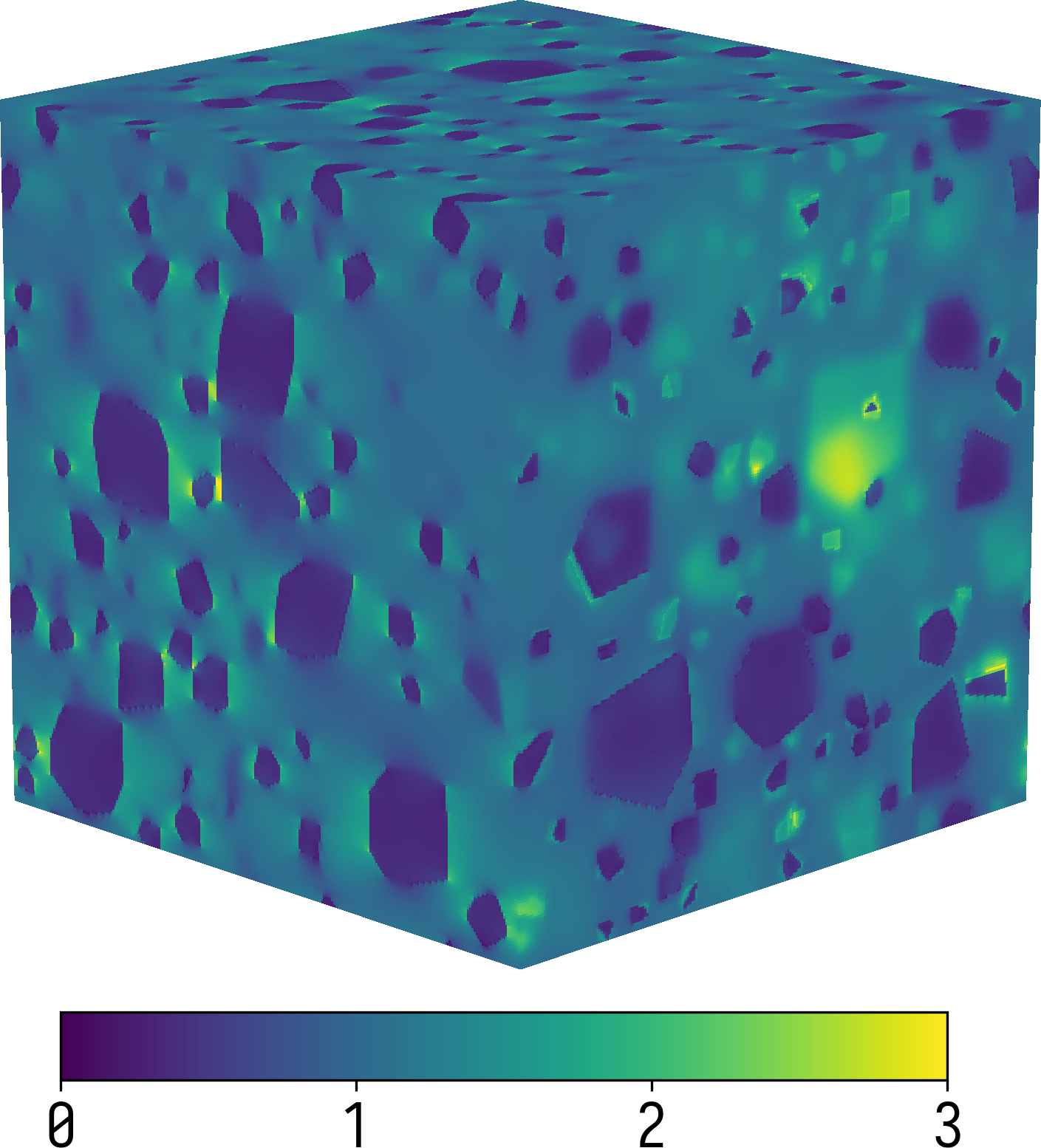}
    \caption{$\mathbb{A}_{1111}^{\mathrm{target}}$}
    \label{fig:final-output-target}
  \end{subfigure}
  \hfill
  \begin{subfigure}[b]{0.32\textwidth}
    \includegraphics[width=\textwidth]{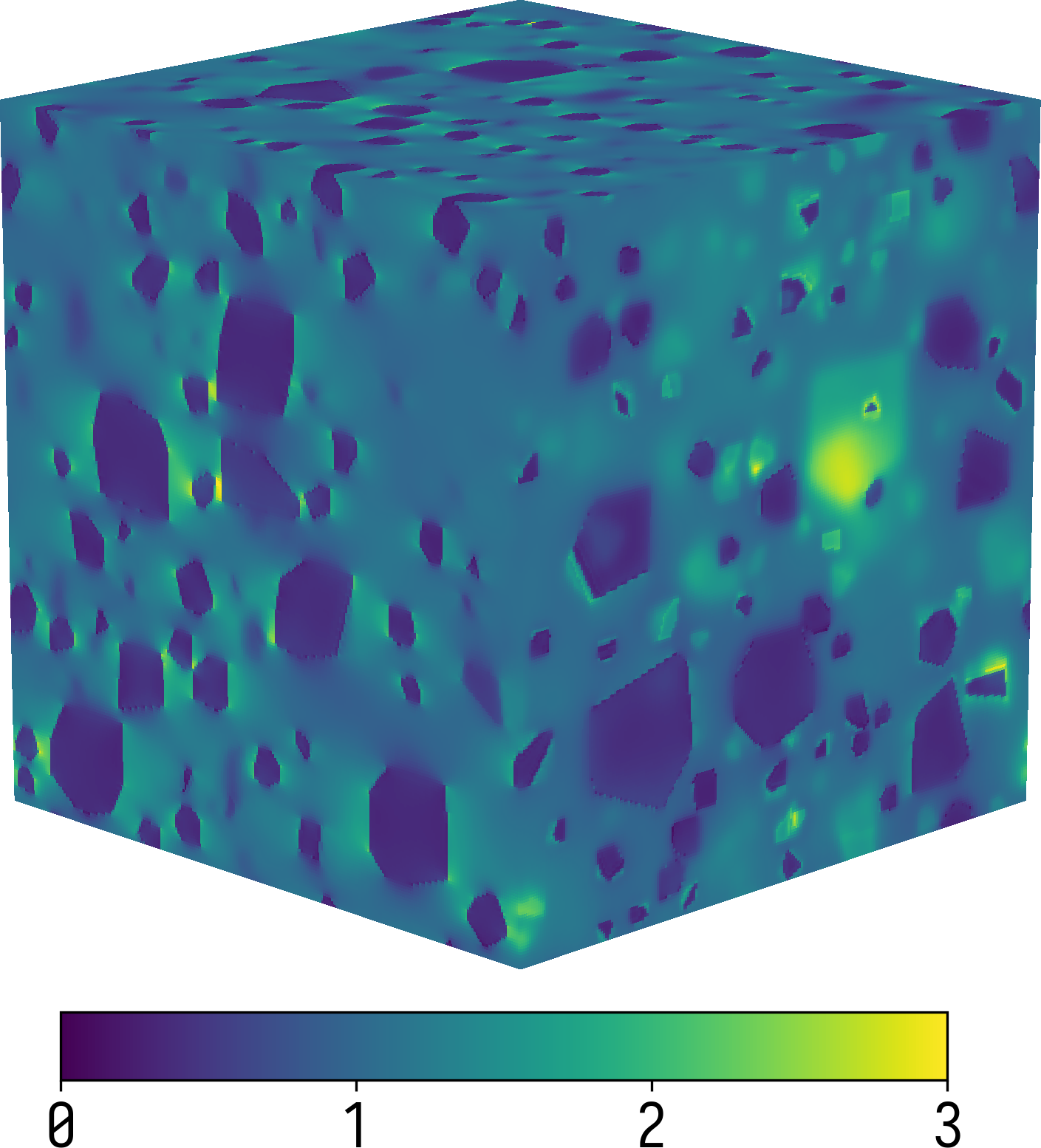}
    \caption{$\mathbb{A}_{1111}^{\mathrm{prediction}}$}
    \label{fig:final-output-prediction}
  \end{subfigure}
  \hfill
  \begin{subfigure}[b]{0.32\textwidth}
    \includegraphics[width=\textwidth]{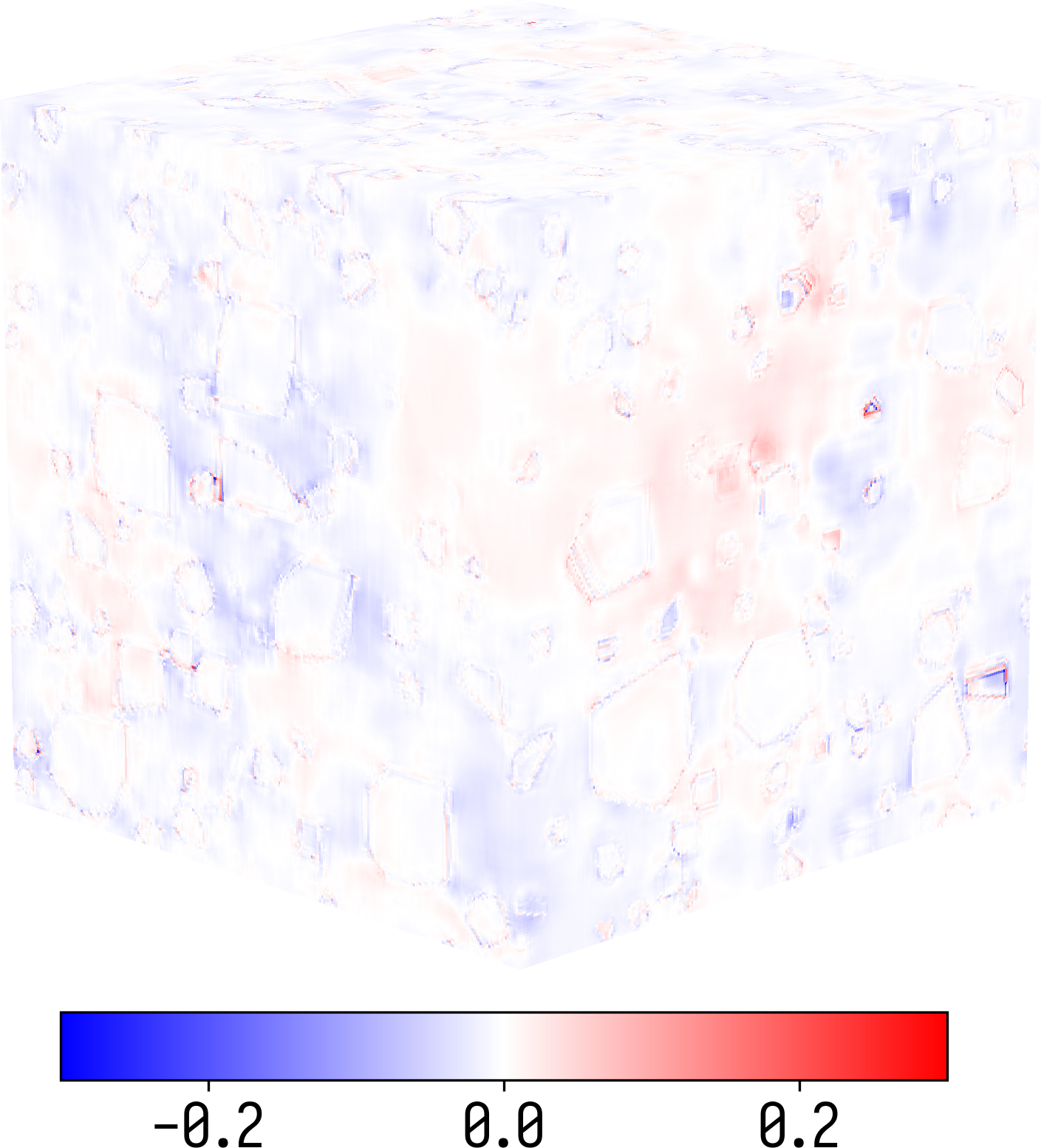}
    \caption{$\mathbb{A}_{1111}^{\mathrm{target}} - \mathbb{A}_{1111}^{\mathrm{prediction}}$}
    \label{fig:final-output-diff}
  \end{subfigure}
  \caption{Validation sample comparison. (a) Target strain localization field (considered as ``ground truth''), (b) Neural network prediction, and (c) Difference between the two.}
  \label{fig:final-output}
\end{figure}

A qualitative assessment of the model's performance is provided in \autoref{fig:final-output}, which compares the first component of the target strain localization field $\mathbb{A}_{1111}^{\mathrm{target}}$ against the neural network prediction $\mathbb{A}_{1111}^{\mathrm{prediction}}$ for a representative validation sample. Visually, the target and prediction maps appear highly similar, capturing the complex heterogeneity of the strain field. The difference map highlights the discrepancy between the two, revealing that the highest errors are concentrated at the interfaces between the aggregates and the cement matrix. This is physically consistent, as these regions exhibit the steepest gradients in mechanical properties and strain concentrations, making them the most challenging features to resolve. Because the reported MAE is averaged over all voxels and components, these local discrepancies are not visible in that scalar: errors in the interfacial regions are larger than the global average, which should therefore not be read as a per-voxel accuracy guarantee.

\FloatBarrier

The scalar MAE of \autoref{eq:loss} averages over all \num{36} components. To verify that the aggregate value does not conceal localized inaccuracies on individual components, \autoref{eq:mae-components} reports the test MAE separately for each component, computed on the held-out test set and arranged in the same $6 \times 6$ Voigt-Mandel form as \autoref{fig:localization}. Each entry \(M^{\mathrm{MAE}}_{ij}\) is the test MAE (in percent) for the component \(\mathbb{A}_{ij}\), where the row index \(i\) denotes the microscopic strain component and the column index \(j\) the macroscopic loading component, both ordered as \((11,22,33,23,13,12)\):
\begin{equation}
  \label{eq:mae-components}
  \mathbf{M}^{\mathrm{MAE}} =
  \begin{bmatrix}
    1.47 & 0.93 & 0.92 & 0.99 & 1.21 & 1.24 \\
    0.90 & 1.50 & 0.93 & 1.23 & 0.96 & 1.18 \\
    0.92 & 0.94 & 1.54 & 1.25 & 1.19 & 1.00 \\
    0.87 & 1.04 & 1.04 & 1.47 & 1.12 & 1.09 \\
    1.07 & 0.89 & 1.04 & 1.09 & 1.47 & 1.11 \\
    1.06 & 1.06 & 0.88 & 1.09 & 1.11 & 1.45
  \end{bmatrix}
\end{equation}
All component errors lie between \SI{0.87}{\percent} and \SI{1.54}{\percent}, with the largest error on the normal $\mathbb{A}_{3333}$ component and the largest coupling error on the $\mathbb{A}_{3323}$ term (\SI{1.25}{\percent}). These component-level values are consistent with the overall test MAE of \SI{1.1184}{\percent} reported above, so the aggregate value does not conceal any component-level outlier.

Finally, the computational efficiency of the trained surrogate model is compared with that of the FFT-based solver in \autoref{tab:final-time}. Both methods were run on an Nvidia H100 GPU in single precision (float32), with a batch size of one and PyTorch JIT compilation. The FFT timing corresponds to the complete localization tensor, i.e., the six unit loadings of \autoref{sec:data-out}, so that both methods produce the full \num{36}-component field. Under these conditions, the solver required \SI{1.39}{\second} per sample at a mean power consumption of \SI{450}{\watt}, whereas a single forward pass of the network completed in \SI{0.16}{\second}, corresponding to a speedup of approximately \num{8.7}. Although the network draws more instantaneous power (\SI{600}{\watt}), likely because the tensor cores are heavily used during inference, its shorter runtime lowers the energy cost per prediction from \SI{625}{\joule} to \SI{96}{\joule}, a reduction by a factor of approximately \num{6.5}. This acceleration confirms the suitability of the proposed surrogate modeling approach for multiscale simulations where numerous microstructural computations are required. This comparison is per prediction and does not include the one-off costs of dataset generation and hyperparameter optimization; the surrogate therefore becomes advantageous only when the number of RVE evaluations is large, as in the coupled multiscale simulations targeted here.

\begin{table}[ht]
  \centering
  \small
  \caption{Comparison of computational cost and power consumption between the FFT-based solver and the trained neural network. Both methods use float32, a batch size of one, and PyTorch JIT compilation on an Nvidia H100 GPU.}
  \label{tab:final-time}
  \begin{tabular}{lcccc}
    \toprule
    \textbf{Hardware} & \textbf{Method} & \textbf{Power consumption} & \textbf{Computation time} & \textbf{Energy per sample} \\
    \midrule
    GPU H100   & FFT-based solver & \SI{450}{\watt} & \SI{1.39}{\second} & \SI{625}{\joule} \\
    GPU H100   & Neural network & \SI{600}{\watt} & \SI{0.16}{\second} & \SI{96}{\joule} \\
    \bottomrule
  \end{tabular}
\end{table}
\FloatBarrier

While speed and accuracy are essential, practical deployment also requires an understanding of how the surrogate generalizes beyond its training distribution.

\subsection{Generalization}
\label{sec:generalization}

Despite their computational advantages, neural network surrogates are inherently constrained by their training distribution: they are reliable only when applied to microstructures that are statistically similar to those seen during training. To quantify this limitation in the context of concrete mesostructures, we assess whether a U-Net trained on one morphology can generalize to unseen aggregate volume fractions and aspect ratios. In addition to direct error measurements, we introduce a Maximum Mean Discrepancy (MMD) metric, computed on the model's internal feature maps, which provides a diagnostic tool to decide whether a new microstructure lies within the training distribution without requiring the FFT reference computation.

To assess the robustness of the proposed framework against morphological variations, a cross-evaluation study was conducted using the optimal hyperparameters identified in \autoref{sec:tuning}. Five distinct neural network instances were trained from scratch, categorized into two groups based on the diversity of their training data. The first group comprised four specialized models (A--D), each trained exclusively on a single morphological configuration defined by specific aggregate volume fractions and aspect ratios (see \autoref{fig:morphologies}). Models A and B corresponded to a \SI{20}{\percent} volume fraction with aspect ratios of \SI{100}{\percent} and \SI{150}{\percent}, respectively, while Models C and D corresponded to a \SI{40}{\percent} volume fraction with the same aspect ratios. The second group consisted of a single generalist model (Model E), trained on the aggregated dataset encompassing all four morphological configurations.

All generalization experiments were conducted at a resolution of $64^3$ voxels. This lower resolution was chosen to enable the rapid training of multiple independent models from scratch while preserving the essential morphological features required to assess cross-configuration generalization. All five models were trained with identical hyperparameters, using a compromise configuration adapted to $64^3$ resolution.

\subsubsection{Mean absolute error}
\label{sec:mae}

To evaluate generalization capabilities, all five models were subsequently tested against the full suite of four morphologies. This experimental design facilitated a direct comparison between in-distribution performance and out-of-distribution generalization. Specifically, it quantifies the performance degradation incurred when a specialized model (e.g., Model A) is applied to unseen microstructural geometries (e.g., the morphology of Model D), relative to the baseline performance of the generalist Model E across all configurations.

\begin{figure}[ht]
  \centering
  \includegraphics[width=0.8\textwidth]{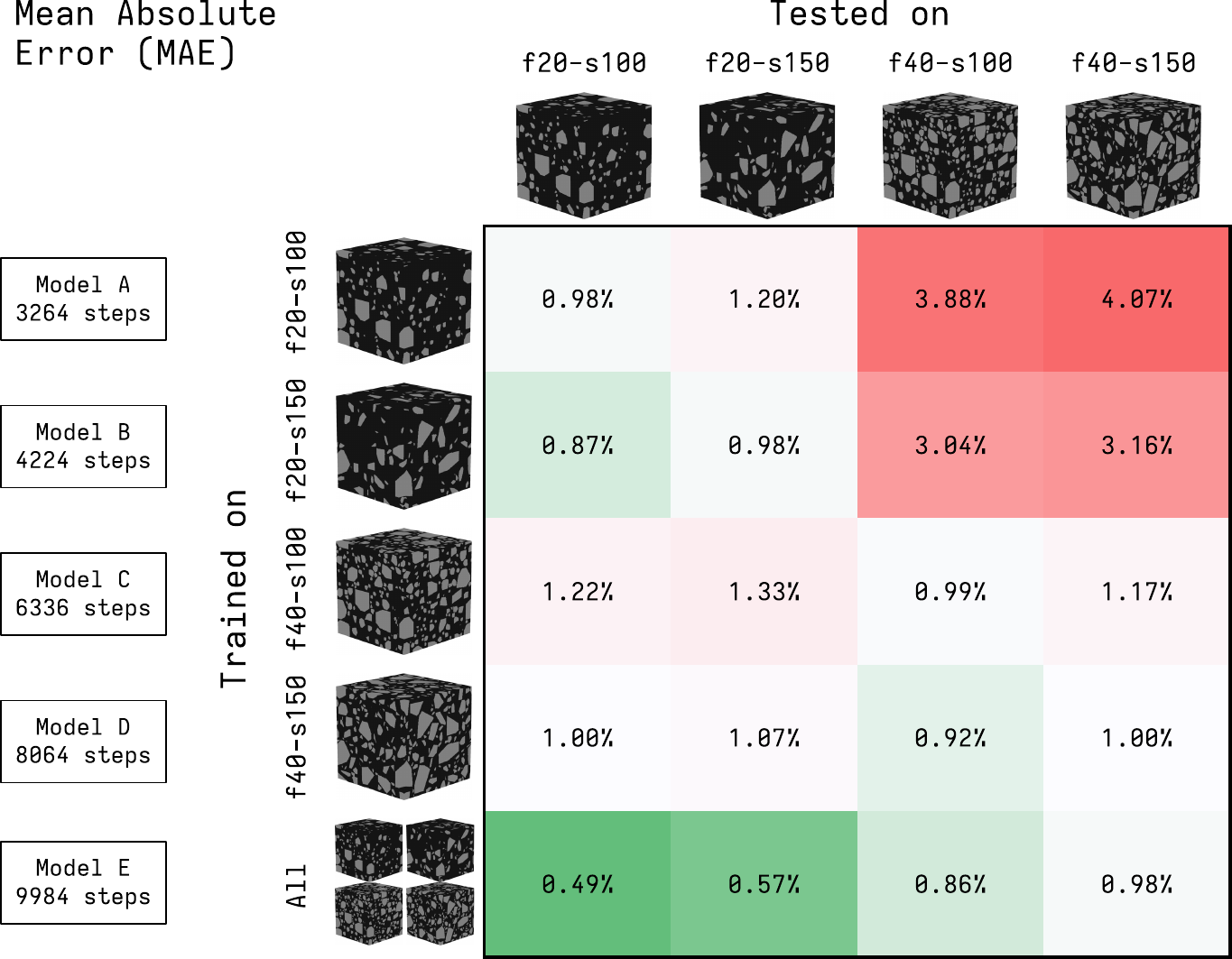}
  \caption{Generalization performance matrix showing Mean Absolute Error (MAE) for specialized and generalist models. Rows denote training configurations and columns denote validation configurations. Training was terminated upon reaching a \SI{1}{\percent} MAE threshold on the training set (step counts indicated on the left). Specialized models display significant error escalation when tested on morphologies with higher volume fractions than those seen during training. In contrast, the generalist model (\texttt{All}), trained on the aggregated dataset, exhibits consistent high accuracy across all morphological variations, outperforming specialized models in out-of-distribution scenarios.}
  \label{fig:generalization-table-mae}
\end{figure}

The five models described above were trained with a convergence criterion set at a Mean Absolute Error (MAE) of \SI{1}{\percent}. As indicated by the step counts in \autoref{fig:generalization-table-mae}, the computational cost to reach this threshold varied significantly by morphology, ranging from \num{3264} steps for the \texttt{f20-s100} configuration to \num{9984} steps for the aggregated dataset. Specialized models exhibit strong in-distribution performance, with validation errors on their respective training morphologies aligning closely with the \SI{1}{\percent} stopping threshold. However, out-of-distribution testing reveals significant limitations in generalization, characterized by a marked asymmetry. Models trained on low volume fraction morphologies ($\mathrm{VF}=\SI{20}{\percent}$) show performance degradation when applied to high volume fraction configurations ($\mathrm{VF}=\SI{40}{\percent}$), with errors ranging between \SI{3.07}{\percent} and \SI{4.43}{\percent}. Conversely, models trained on $\mathrm{VF}=\SI{40}{\percent}$ data generalize more effectively to $\mathrm{VF}=\SI{20}{\percent}$ validation cases, maintaining errors below \SI{1.5}{\percent}. The generalist model (labeled \texttt{All}), trained on the full dataset, demonstrates superior robustness. Despite requiring the longest training duration, it achieves the lowest error rates across all validation configurations, consistently maintaining an MAE below \SI{1}{\percent} (ranging from \SI{0.52}{\percent} to \SI{0.98}{\percent}).

\FloatBarrier

\subsubsection{Maximum mean discrepancy}
\label{sec:mmd}

Maximum Mean Discrepancy (MMD), a kernel-based two-sample statistic widely used for distribution-shift quantification \cite{gretton2012}, has been employed in microstructure-informed deep learning to assess surrogate prediction reliability by quantifying distributional shifts in learned feature representations \cite{gupta2026}. Following this approach, MMD is computed in the learned feature space of each U-Net model to quantify the discrepancy between validation microstructures and the training distribution.

Features are extracted from the \texttt{encoder\_3\_conv\_0} layer, which provides a deep representation of morphology-relevant and prediction-relevant features. MMD computed from this representation consistently exhibits a strong correlation with prediction error.

MMD is evaluated using a Radial Basis Function (RBF) kernel, with the kernel bandwidth determined from the reference features using the median pairwise-distance heuristic. For each specialized model, the \num{192} training microstructures are divided into \num{160} reference samples and \num{32} held-out in-distribution samples. The reference samples define the latent training distribution, while the held-out samples establish an independent in-distribution baseline MMD. Each validation microstructure is compared individually against the reference distribution, and the MMD values are averaged over the \num{32} microstructures in each validation dataset. For the generalist model, the reference distribution consists of \num{160} samples from each of the four training classes (\num{640} samples total), and the baseline is obtained by averaging the class-specific baseline MMDs.

The distribution shift is expressed relative to the in-distribution baseline as
\begin{equation}
  \mathrm{MMD}_{\mathrm{rel}} = \frac{\mathrm{MMD}_{\mathrm{val}} - \mathrm{MMD}_{\mathrm{baseline}}}{\mathrm{MMD}_{\mathrm{baseline}}}.
\end{equation}

A relative MMD near zero indicates a distribution comparable to the in-distribution baseline, while increasing positive values indicate greater departure from the training distribution. Negative values indicate a discrepancy below the estimated in-distribution baseline and are therefore also interpreted as being within the baseline distributional variability. Because each surrogate learns a distinct latent representation and kernel bandwidth, relative MMD values are interpreted primarily within each model rather than quantitatively across models.

\begin{figure}[ht]
  \centering
  \includegraphics[width=0.8\textwidth]{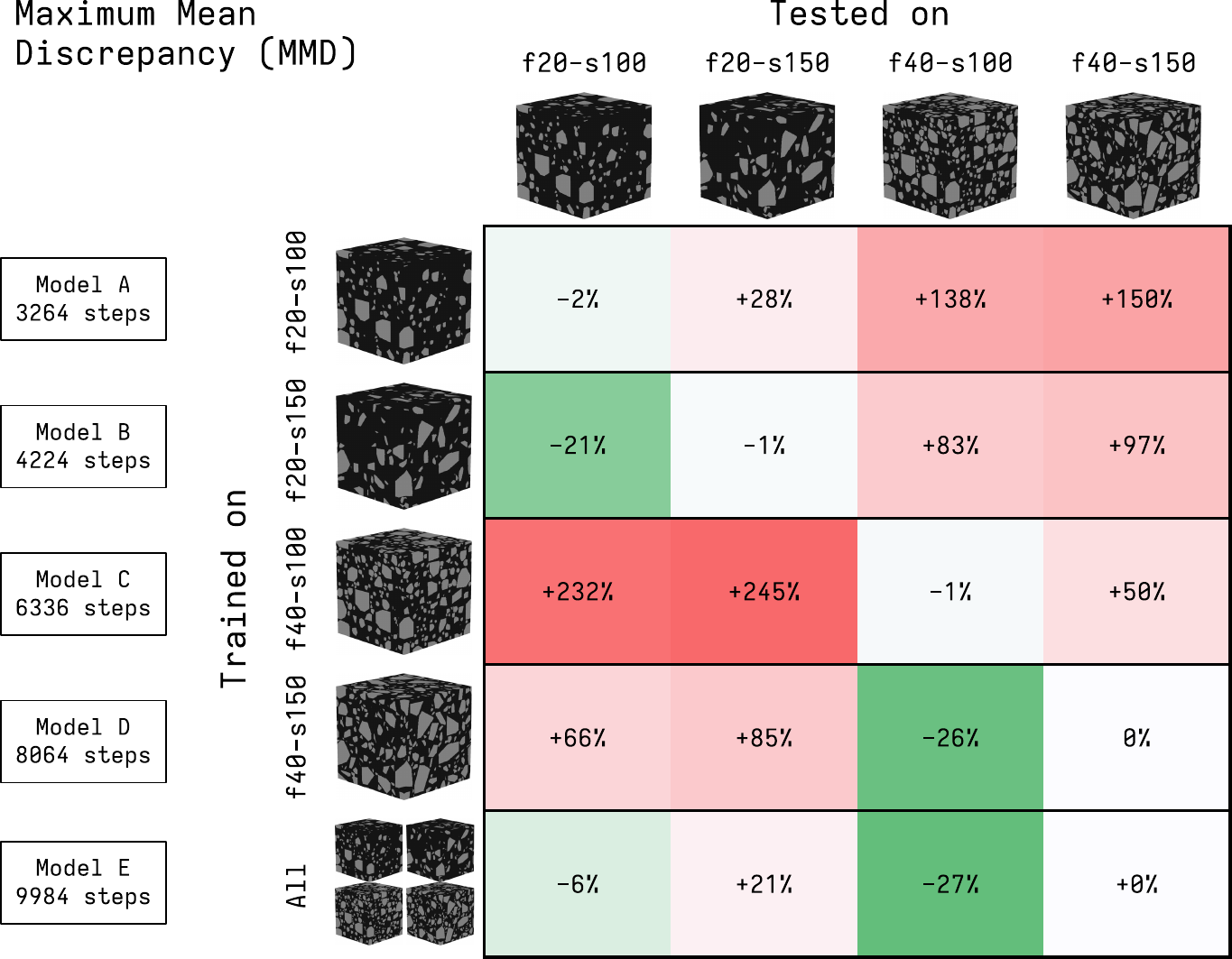}
  \caption{Mean relative Maximum Mean Discrepancy (MMD) between the latent representations of validation microstructures and the corresponding training reference distributions. Values represent the percentage change relative to the in-distribution baseline MMD for each model. Each row should be interpreted independently because every model learns a distinct latent representation. Green and red shading correspond to small and large latent distribution shifts, respectively. The encoder is more sensitive to aggregate volume fraction than to aggregate shape, and the generalist model (\texttt{All}) exhibits uniformly low MMD across all validation configurations.}
  \label{fig:generalization-table-mmd}
\end{figure}

\autoref{fig:generalization-table-mmd} reports the relative MMD, expressed as the percentage change with respect to each model's in-distribution baseline MMD. For each specialized model, the baseline MMD was calculated using a held-out subset of the training data. The validation microstructures were then compared against the corresponding training distribution. Since each model learns a distinct latent representation, the MMD values should be interpreted primarily within rows rather than compared quantitatively across models.

The specialized models exhibit progressively larger relative MMD as the validation microstructures deviate from their corresponding training morphology. Changes in aggregate volume fraction consistently produce substantially larger latent distribution shifts than changes in aggregate shape. In contrast, the generalist model, trained on the combined dataset, exhibits small relative MMD variations across all four microstructure classes, indicating broader coverage of the microstructure design space.

Although MMD does not fully explain why Models~C and~D generalize relatively well to lower-volume-fraction microstructures, a large MMD reliably signals out-of-distribution inputs and thus serves as a practical, computationally inexpensive indicator of surrogate trustworthiness.

\FloatBarrier

\autoref{tab:correlation} reports the Pearson correlation coefficient ($r$) between the relative MMD and the corresponding prediction error (MAE) across the four validation datasets for each specialized model. The consistently high positive correlations indicate that increasing latent distribution shift is associated with increasing prediction error, supporting MMD as an effective indicator of distribution shift, model generalization, and prediction reliability for previously unseen microstructure datasets. A Pearson correlation is not reported for the generalist model because all evaluation datasets are represented in its training distribution, and the remaining prediction errors are no longer primarily governed by distribution shift.

\begin{table}[ht]
  \centering
  \caption{Pearson correlation coefficient representing the linear correlation between the MAE presented in \autoref{fig:generalization-table-mae} and the MMD presented in \autoref{fig:generalization-table-mmd}.}
  \label{tab:correlation}
  \begin{tabular}{lc}
    \toprule
    \textbf{Model} & \textbf{Correlation $r$} \\
    \midrule
    Model A & $0.98$ \\
    Model B & $0.99$ \\
    Model C & $0.88$ \\
    Model D & $0.86$ \\
    \bottomrule
  \end{tabular}
\end{table}
\FloatBarrier

\section{Discussion}
\label{sec:discussion}

Beyond the quantitative results, several design and deployment implications emerge from this study.

A key contribution of this work is the comprehensive hyperparameter optimization framework, which identified robust design rules for adapting the U-Net architecture to increasing problem complexity. Our investigation has revealed that most architectural and training hyperparameters can be fixed across resolutions, output dimensionalities, and morphological diversities, with only two parameters requiring careful scaling: the number of channels, which controls the accuracy-speed trade-off, and the number of levels, which should be increased with resolution to maintain a minimal encoder feature-map resolution of $4^3$. Secondary design choices, including $3^3$ non-separated convolutions, direct skip connections, omitted normalization and bias terms, and the SiLU activation function, were found to yield the best compromise between accuracy and computational efficiency. Notably, maintaining a constant channel depth across encoder levels, rather than doubling the number of channels after each downsampling step as is conventional in most U-Net implementations, was found to deliver comparable predictive accuracy while significantly reducing memory consumption. This departure from the standard U-Net convention is particularly important for 3D microstructures, where memory constraints rapidly become the limiting factor.

The choice of optimizer also proved influential, with the Muon algorithm consistently outperforming AdamW and SGD with momentum for this high-dimensional image-to-image regression task. To the best of our knowledge, this work constitutes the first application of the Muon optimizer to a continuum solid mechanics problem. Muon has recently become the de facto optimizer for the pretraining of several frontier large language models \cite{jordan2024,liu2025}, where the Newton-Schulz orthogonalization of the update matrix is valued for its training stability and efficiency. The present results suggest that this stabilization is not specific to language modeling, but extends to the high-dimensional, spatially structured tensor-field regression problems encountered in computational homogenization.

The most important practical takeaway is that the ``optimal'' model configuration is not universal: it critically depends on the user's objectives. The dual-criterion selection strategy proposed in this study distinguishes between accuracy-optimal and speed-optimal configurations. Accuracy-optimal models achieve validation errors below \SI{0.4}{\percent} but require larger networks with inference times exceeding \SI{5}{\milli\second} per sample, whereas speed-optimal models reduce inference time by up to a factor of six while retaining errors below the \SI{1}{\percent} threshold. This underscores the necessity of clearly defining application-specific goals before hyperparameter optimization: a surrogate intended for offline homogenization of a limited set of microstructures may prioritize accuracy, whereas one embedded in a coupled multiscale simulation with millions of RVE evaluations may prioritize throughput.

Closely related to this objective-dependent optimum is the influence of the stopping criterion. The hyperparameter search employed time-constrained training budgets, and the final model was trained until validation error stagnated. Different training durations or convergence thresholds would naturally shift the accuracy-speed trade-off analysis. For instance, shorter budgets favor smaller, faster-converging networks, while extended training can unlock additional accuracy from larger architectures. Therefore, the reported optimal hyperparameters should be understood in the context of the specific computational budget and error tolerance under which they were obtained. Researchers and practitioners should align the optimization objective and stopping criterion with the operational constraints of their target multiscale framework.

Generalization experiments revealed important limitations inherent to data-driven surrogates. Specialized models trained on a single morphology exhibited strong in-distribution performance but significant degradation when applied to microstructures with higher aggregate volume fractions than those seen during training. A generalist model trained on a diverse ensemble of morphologies consistently outperformed specialized models, maintaining validation errors below \SI{1}{\percent} across all configurations. This highlights the value of training-data diversity when the surrogate is expected to handle microstructural variability. To mitigate reliability concerns outside the training distribution, we proposed the use of a Maximum Mean Discrepancy metric computed on internal feature maps as a diagnostic indicator. Such a metric can flag out-of-distribution inputs without requiring costly reference computations, thereby enabling informed decisions about surrogate trustworthiness in operational settings.

While this study establishes a robust framework for linear elastic regimes, several assumptions present fertile ground for future investigation. First, the current methodology is restricted to linear elasticity; extending it to inelastic behavior, such as plasticity or damage evolution, would require accounting for loading history. One route is to treat the loading sequence as an additional input dimension, yielding a 4D input domain (3D structure + 1D loading sequence). Although computationally demanding, recent advances in sequence modeling for mechanics \cite{saha2024} suggest that this route is conceptually feasible. Second, the scalability of the proposed architecture warrants further exploration regarding domain size transferability. Specifically, future work could investigate whether a U-Net trained on smaller representative volume elements can reliably predict fields in significantly larger domains, potentially offering a strategy to reduce the computational burden associated with generating high-resolution training data.

\section{Conclusion}
\label{sec:conclusion}

This study has presented a methodology for developing U-Net surrogate models to predict the full \num{36}-component strain localization tensor field in 3D voxelized concrete mesostructures. By framing the mechanical problem in terms of the localization tensor, the trained surrogate enables the reconstruction of microscopic strain and stress fields for arbitrary macroscopic loadings, offering an efficient alternative to traditional solvers. The hyperparameter study further showed that only network width and depth need to be scaled with problem complexity, while most other design choices can remain fixed.

The final trained surrogate achieved a test MAE of \SI{1.118}{\percent}, essentially identical to the validation MAE, and an inference speedup of approximately $8.7\times$ for the full \num{36}-component tensor compared to the FFT-based solver on an Nvidia H100 GPU, with a corresponding reduction in total energy consumption. The successful scaling of the U-Net surrogate to high-resolution three-dimensional microstructures represents the principal novelty of this work. Extending deep-learning-based field prediction to $256^3$ voxel domains is challenging because of the substantial memory requirements associated with volumetric feature maps and the computational burden of regressing high-dimensional tensor fields. Nevertheless, the optimized architecture and training pipeline proved capable of handling these large 3D volumes, delivering predictions faster than an FFT-based solver, itself already recognized as a highly efficient numerical method for periodic heterogeneous media. Qualitative assessment confirmed that the largest prediction errors are concentrated at aggregate-matrix interfaces, which is physically consistent with the steep property gradients and strain concentrations present in these regions.

Taken together, this work demonstrates that carefully optimized U-Net surrogates can deliver accurate and computationally efficient predictions of strain localization fields in heterogeneous materials, provided that the model configuration is tailored to the specific accuracy, speed, and generalization requirements of the intended application. The clear definition of optimization objectives and stopping criteria is essential to navigating the inherent trade-offs and unlocking the full potential of machine learning surrogates in multiscale mechanical simulations.

\section*{Acknowledgments}

The authors gratefully acknowledge the financial support of the Swiss Federal Office of Energy (Contract No. sl/502787-01) and the Swiss National Science Foundation (Grant No. 221743). We also thank Somdatta Goswami and Nicolas Richart for their helpful discussions and constructive feedback.

\footnotesize
\bibliographystyle{ieeetr}
\bibliography{ref.bib}

\clearpage

\normalsize
\appendix

\end{document}